\documentclass[sigconf]{acmart}
\usepackage{amsmath}
\usepackage{subfigure}
\AtBeginDocument{%
  }

\setcopyright{acmlicensed}
\copyrightyear{2024}
\acmYear{2024}
\acmDOI{XXXXXXX.XXXXXXX}

\acmConference[RecSys '24]{Eighteenth ACM Conference on Recommender Systems}{Oct. 14--18,
  2024}{Bari, Italy}
\acmISBN{978-1-4503-XXXX-X/18/06}

\begin{document}

\title{Calibrating the Predictions for Top-N Recommendations}


\author{Masahiro Sato}
\affiliation{%
  \institution{FUJIFILM}
  \city{Akasaka}
  \state{Tokyo}
  \country{Japan}
}
\email{masahiro.a.sato@fujifilm.com}

\renewcommand{\shortauthors}{Sato et al.}

\begin{abstract}
	Well-calibrated predictions of user preferences are essential for many applications.
	Since recommender systems typically select the top-N items for users, calibration for those top-N items, rather than for all items, is important. 
	We show that previous calibration methods result in miscalibrated predictions for the top-N items, despite their excellent calibration performance when evaluated on all items. 
	In this work, we address the miscalibration in the top-N recommended items.
	We first define evaluation metrics for this objective and then propose a generic method to optimize calibration models focusing on the top-N items.
	It groups the top-N items by their ranks and optimizes distinct calibration models for each group with rank-dependent training weights. 
	We verify the effectiveness of the proposed method for both explicit and implicit feedback datasets, using diverse classes of recommender models.
\end{abstract}





\maketitle

\section{Introduction}
Recommender models predict user preferences based on explicit (e.g., ratings) or implicit (e.g., clicks or views) feedbacks and recommend the top-N items for each user.
The predicted values of user preferences are useful for many applications, not only for ranking items~\cite{menon2012predicting, kweon2022obtaining}.
For example, displaying the predicted ratings together with recommended items can help users decide whether to consume the items~\cite{herlocker2000explaining}, especially when the consumption takes substantial time (e.g., watching videos) or money (e.g., online purchases).
Another application is multi-objective recommendations~\cite{ribeiro2012pareto,rodriguez2012multiple}; business owners might want to rerank recommended items, balancing estimated purchase probabilities and profitability of items~\cite{de2024model}.
These applications require that predicted values are well-calibrated.
If a recommender outputs 0.5 as purchase predictions, 5 out of 10 items with such a prediction should be purchased.
Similarly, if ratings of several items are predicted to be 4.6, we expect the average ratings after the consumption of those items to be around 4.6.

It has been known that accurate classification or ranking models can produce poorly-calibrated predictions~\cite{guo2017calibration, kweon2022obtaining}.
Therefore, there has been a wide range of research on obtaining well-calibrated predictions.
For that purpose, a widely used approach is post-hoc calibration: a method that applies a calibration model to map the initial prediction scores of a classifier into estimates of ground truth probabilities.
While most calibration research has focused on classification models~\cite{zadrozny2002transforming, niculescu2005predicting, guo2017calibration}, recent works have investigated calibration for ranking models~\cite{menon2012predicting, kweon2022obtaining}.
These methods have successfully improved calibration performance when measured on the entire data space.

However, as we show in this paper, predictions for top-N recommended items can be miscalibrated even if the predictions are apparently well-calibrated when evaluated on all items.
Calibration on top-N items is vital for recommendations since predicted values of top-N items, rather than those of all items, have subsequent uses in most applications.
Despite its importance, this issue has been mostly overlooked.
Very recent work tackled a relevant problem: theoretical overestimates of predicted values in top-N items~\cite{fan2023calibration}.
In practice, however, miscalibration in top-N items can be both overestimates and underestimates.
Because of the above complexity, the issue remains unsolved.

In this work, we address calibration on top-N recommended items.
We first show that seemingly well-calibrated predictions on all items could be miscalibrated on top-N items (Section 3.2).
To properly evaluate the calibration performance on top-N items, we define evaluation metrics for this purpose (Section 4.1).
Then, we propose a method to optimize calibration models focusing on top-N items (Section 4.2).
It first extracts top-N items and groups them by their ranks, and then constructs calibration models for each ranking group.
To prioritize calibration in higher ranks, we further impose rank-discounting weights when training calibration models.
The proposed method is a generic optimization method and applicable to many existing calibration models.
We verify the effectiveness of our proposed method for both rating prediction and preference probability prediction tasks with various recommender models and calibration models (Section 5).

%[Summary of contributions]
%The contribution of this paper are summarized as follows.
%\begin{itemize}
%	\item We propose evaluation metrics for miss-calibration in top-N items (Section 4.1).
%	\item We present a generic optimization method of calibration models to address the miss-calibration (Section 4.2).
%	\item We demonstrate the effectiveness of our optimization method through experiments of both explicit and implicit feedback with various recommender models (Section 5). 
%\end{itemize}

\section{Related Work}
Calibration models map the initial predictions of a classifier or a ranker into calibrated predictions.
Most calibration models are categorized into two groups: non-parametric and parametric models.
Non-parametric models, which include histogram binning~\cite{zadrozny2001obtaining} and isotonic regression~\cite{zadrozny2002transforming}, divide the input space into multiple bins and find corresponding outputs for each bin.
Parametric models learn specific parametric functions for calibration.
The representative ones are Platt scaling~\cite{platt1999probabilistic} and Beta calibration~\cite{kull2017beta}.
Recent work has studied calibration for ranking models.
Menong et al. applied isotonic regression for predicted scores of ranking models~\cite{menon2012predicting}.
Kwong et al. extended Platt scaling to account for score distributions in ranking models and proposed Gaussian calibration and Gamma calibration~\cite{kweon2022obtaining}.\footnote{Their work also addresses the bias coming from the missing-not-at-random nature of user feedback~\cite{schnabel2016recommendations,saito2020unbiased}. Such a debiasing method is orthogonal to our method. For simplicity, we focus on the miscalibration issue in this paper.}
In this work, we apply our optimization method to these calibration models. 

Very recently, Fan et al. proposed an algorithm to mitigate overestimates in calibrated scores for top-N items~\cite{fan2023calibration}.
It quantifies the overestimates from the variance of predictors and then subtracts them from the calibrated scores.
However, we find that there are also underestimates in top-N items (Fig. \ref{fig:reliability_diagram}).
Our method can prevent both overestimates and underestimates.
We compare our method with their method in the experiment section.

Steck~\cite{steck2018calibrated} proposed a calibration to ensure that a recommendation list has the item category proportion close to that of the user's past interacted items (i.e., if a user has watched 70 romance and 30 action movies in the past, the recommendation list should include 70\% romance and 30\% action movies).
Such a calibration has been extensively studied~\cite{kaya2019comparison,seymen2021constrained,naghiaei2022towards,abdollahpouri2023calibrated} after the above pioneering work.
While these works also calibrate the top-N recommended items, the pursued calibration is different from ours; theirs focused on item category proportion and ours addresses predicted scores.

\section{Background}
\subsection{Preliminary}
Let $\mathcal{U}$ and $\mathcal{I}$ be sets of users and items, respectively.
Let $y_{ui}$  denote the feedback of user $u \in  \mathcal{U}$ on item $i \in  \mathcal{I}$.
The feedback $y_{ui}$ might be either explicit (e.g., five-scale ratings) or implicit (e.g., purchases or views) signals of user preferences.
A recommender model $f: \mathcal{U} \times \mathcal{I} \to \mathbb{R}$ typically predicts the user feedback directly or predicts scores that correlate with the feedback to rank items.
In this work, we consider a post-processing calibration that maps the model output  $s_{ui} = f(u,i)$ to the calibrated prediction $\hat{y}_{ui} = g(s_{ui})$.

In case of implicit binary feedback, the model is well-calibrated if the calibrated output matches the probability of positive feedback (e.g., purchase probability).
The calibration error (CE) is hence measured in terms of the difference between true probability and predicted probability: $CE = |P(y=1| \hat{y}) -  \hat{y}|$. 
While most calibration researches focus on calibration in class probabilities, we also study regression tasks (e.g., rating prediction).
For both cases, CE can be measured by the difference between the expectation of feedback and its prediction: $CE = |\mathbb{E} [y| \hat{y}] -  \hat{y}|$.
To evaluate a model, the expectation of CE (ECE) is estimated empirically by binning samples~\cite{roelofs2022mitigating}.
\begin{equation}
	ECE = \mathbb{E} [CE] \approx \sum^M_{m=1} \frac{|B_m|}{n} \left| \frac{\sum_{k \in B_m} y_k}{|B_m|} - \frac{\sum_{k \in B_m} \hat{y}_k}{|B_m|}  \right|,
\end{equation}
where $B_m$ is $m$-th bin containing samples that have close predictions and $n$ is the total number of samples.
The difference in the means of feedback and prediction is calculated for each bin, and the differences are averaged for all bins.

\subsection{Miscalibration in top-N items}
\begin{figure}[htbp]
	\begin{center}
			\subfigure[Reliability diagram (N=20)]{\includegraphics[width=0.23\textwidth]{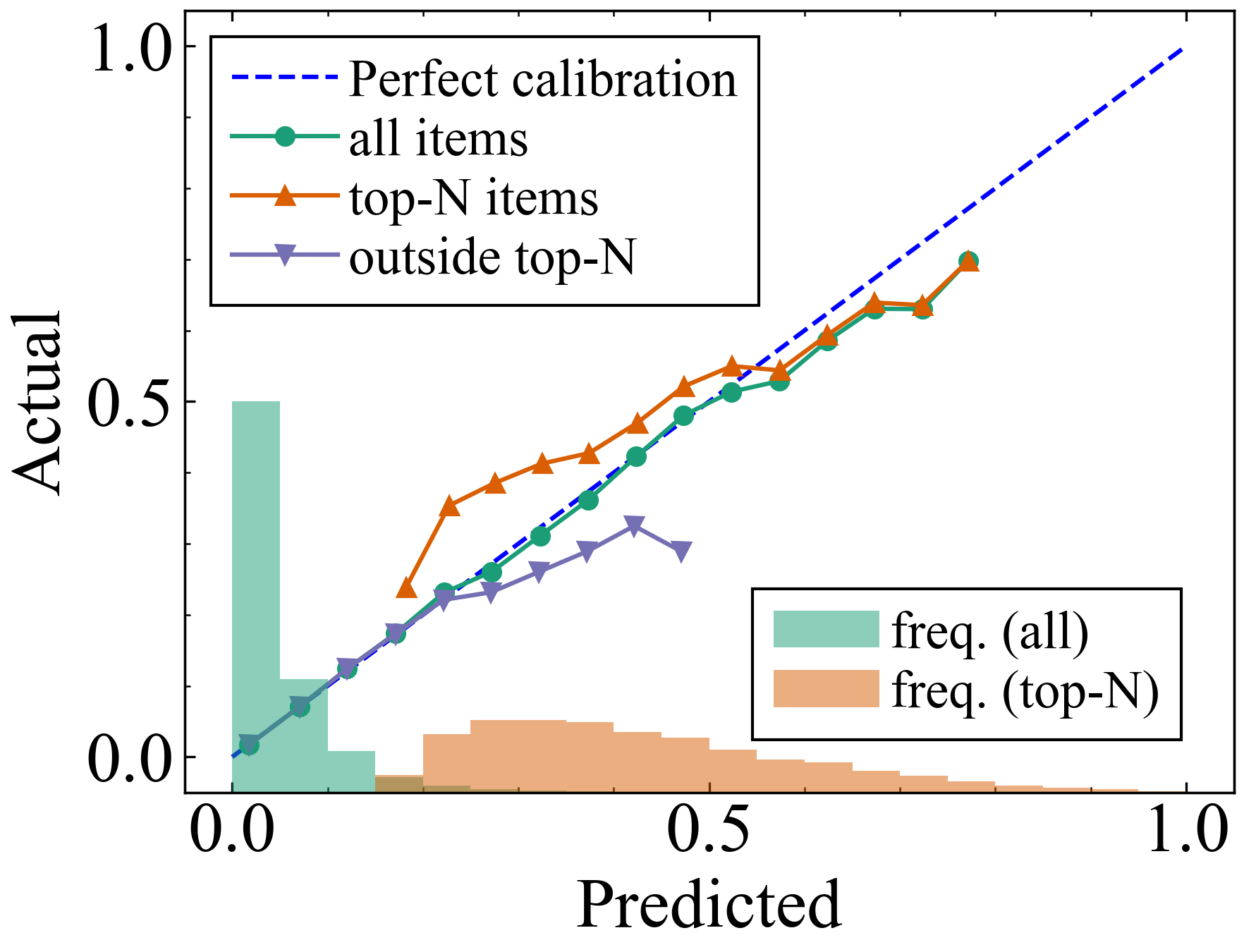}}
			\Description[Line graph showing the reliability diagram for all items, top-N items, and outside top-N items.]{Line graph showing the actual probability of preference from 0.0 to 1.0 on the Y axis against the predicted probability of preference from 0.0 to 1.0 on the X axis. The plots for the all items are close to the perfect calibration line up to 0.5 on the X axis, and they deviate toward downside over 0.5. From 0.25 to 0.5, the plots of top-N items are over the all items and the plots of outside top-N items are under the all items. The relative frequency of the predicted values for all items and top-N items are overlayed on the figure. The all items distribute mostly below 0.2, while the top-N items distribute from 0.2 to 0.7 on the X axis.}
			\subfigure[Rank-based calibration plot]{\includegraphics[width=0.23\textwidth]{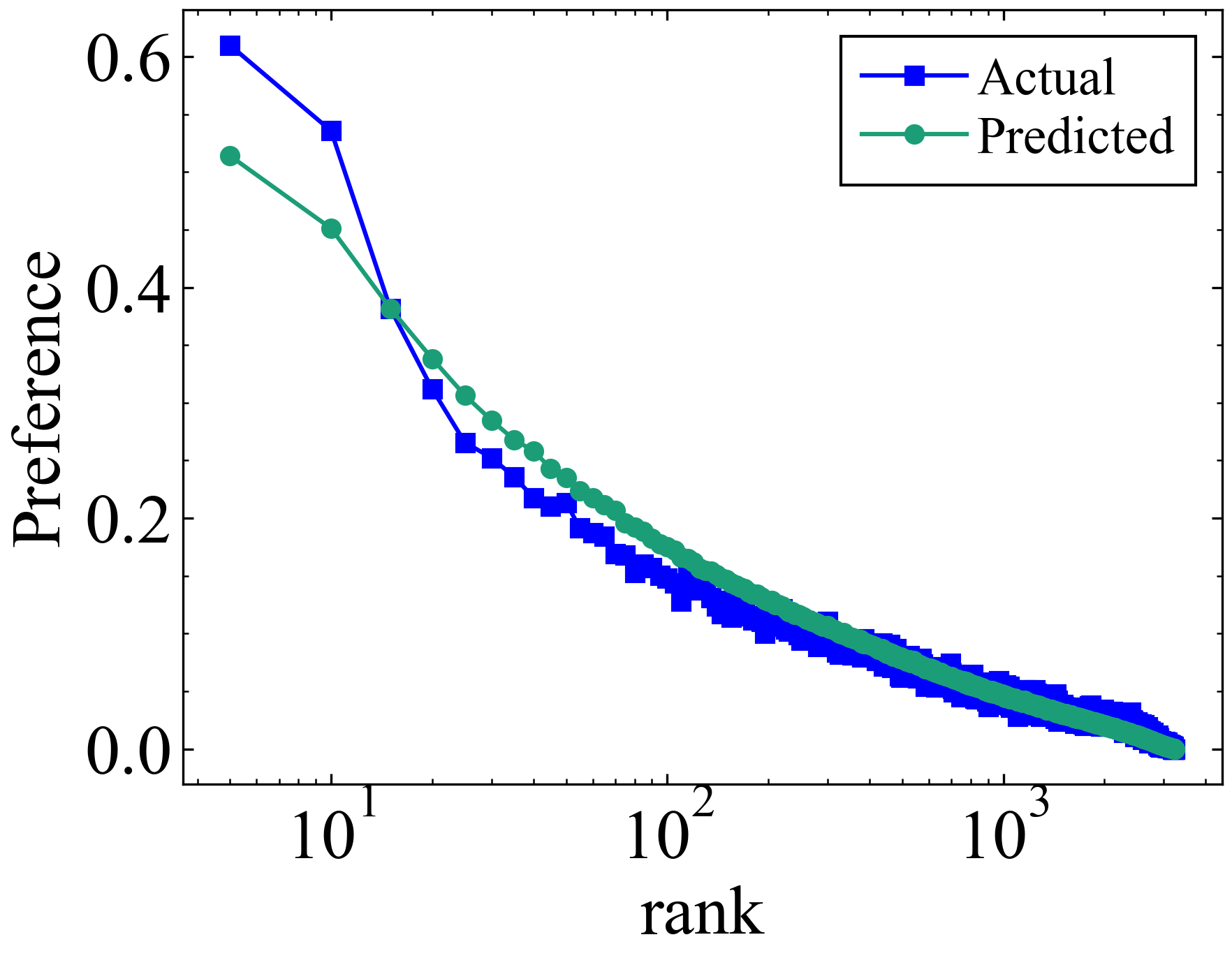}}
			\Description[Line graph showing the preference probabilities of actual and predicted at varied ranking positions.]{Line graph showing the preference probabilities from 0.0 to 0.6 on the Y axis against the rank positions from 5 to 3000 in the logarithmic scale on the X axis. The both plots for the actual and predicted probabilities monotonically decrease as the rank position increases. The actual probabilities are over the predicted until rank = 10. From rank = 20 to rank = 200,  The predicted probabilities are over the actual probabilities. }
			\caption{Calibration plots for NCF with Gaussian calibration applied to preference prediction in the Kuairec dataset.}
			\label{fig:reliability_diagram}
		\end{center}
\end{figure}

Previous works aim to calibrate predictions for all samples (i.e., all user-item pairs in the case of recommendations).
Fig. \ref{fig:reliability_diagram} (a) shows the reliability diagram~\cite{degroot1983comparison,niculescu2005predicting} of calibrated predictions for user preference probabilities.\footnote{The details of the experimental setting are described later in the experiment section.}
It plots the means of predictions (horizontal axis) and observed feedbacks (vertical axis).
The dotted line shows the perfect match between predictions and truths.
The calibration model was trained for all items in a separate validation dataset, as is common in previous works~\cite{menon2012predicting, kweon2022obtaining}.
The reliability diagram calculated for all items is mostly close to the perfect line, meaning that the predictions are fairly well-calibrated for all items.

However, there is a heavy miscalibration for top-N items in the same calibration model.
In the high probability region, the plots for top-N items are below the perfect line, meaning that the predicted values are larger than the actual values.
In the middle probability region, on the contrary, predictions are smaller than truths.
Despite such miscalibration, ECE is 0.001, which implies a low error for predictions in [0, 1] range.
This discrepancy comes from two reasons.
Firstly, most items have predictions around [0.0, 0.1] and the metric is insensitive to miscalibration in the scarce high probability region.
Secondly, in the middle probability region, top-N and outside top-N items have miscalibration in opposite directions, which are canceled out when ECE is calculated for all items.
Further, the miscalibration is rank-dependent.
Fig. \ref{fig:reliability_diagram} (b) shows the means of predictions and ground truths at different ranking positions (grouped by 5 ranks).
While low-ranked items (rank > 200) are well-calibrated, the clear miscalibration appears at highly ranked items. 
Directions (i.e., overestimates or underestimates) of the miscalibration and their degrees vary at different ranks.
The results in Fig. \ref{fig:reliability_diagram} (a) and (b) demonstrate that 1) previous metric ECE cannot quantify the miscalibration in top-N items, and 2) previous calibration methods are not suitable for calibrating top-N items.

\section{Proposed Methods}
\subsection{Evaluation Metrics}
To quantify the calibration in top-N items, we define ECE only for top-N items.
\begin{equation}
	ECE@N = \sum^{M'}_{m=1} \frac{|B'_m|}{n'} \left| \frac{\sum_{k \in B'_m} y_k}{|B'_m|} - \frac{\sum_{k \in B'_m} \hat{y}_k}{|B'_m|}  \right|,
\end{equation}
where $n'$ is the total number of samples ranked within top-N and $B'_m$ is $m$-th bin among those top-N items.
While conventional ECE for calibration models in Fig. \ref{fig:reliability_diagram} (a) is 0.001, ECE@N is 0.070 for the same models, reflecting the miscalibration in top-N items. 

Considering that higher-ranked items have larger importance, we further propose rank-discounted ECE (RDECE).
\begin{equation}
	RDECE@N = \frac{N}{\sum_{r}  w_r} \sum^N_{r=1} \frac{w_r |B'_r|}{n'} \left| \frac{\sum_{k \in B'_r} y_k}{|B'_r|} - \frac{\sum_{k \in B'_r} \hat{y}_k}{|B'_r|}  \right|,
\end{equation}
where $w_r$ is a discounting weight that decreases with the rank $r$ of item $k$ and $B'_r$ is $r$-th bin that contains items ranked in $r$.
In this paper, we use $w_r = 1/r$ that is a popular discount used in the mean reciprocal rank~\cite{voorhees1999trec,zangerle2022evaluating}, for example.
The formulation of RDECE is inspired by variable-based ECE (VECE)~\cite{kelly2023variable}, which measures systematic miscalibration depending on a variable of interest.
RDECE@N can be regarded as an extension of VECE with variable weights, and the variable is the item rank.

\subsection{Calibration Method}
\begin{figure}[htbp]
	\centering
	\includegraphics[width=1.0\linewidth]{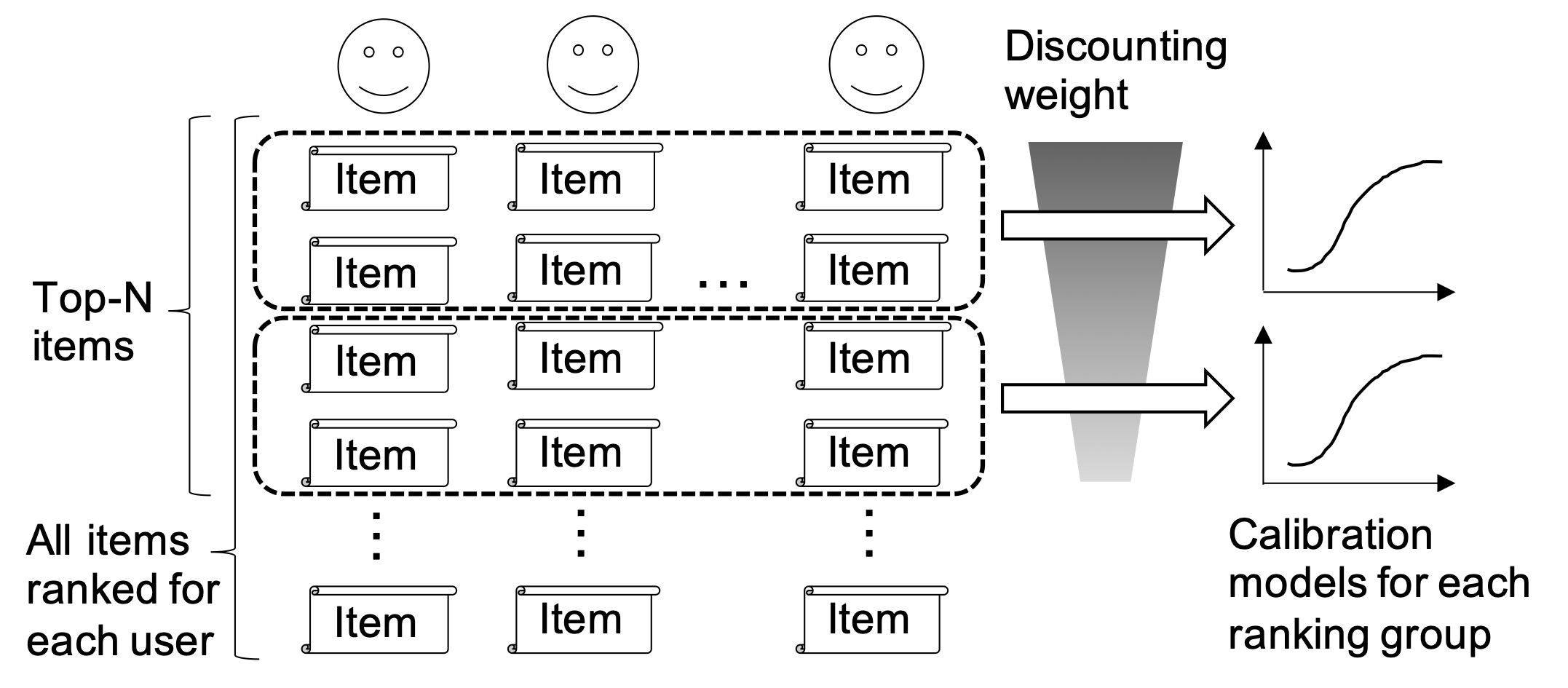}
	\caption{Proposed calibration method. Top-N items are grouped by their ranks. Then calibration models for each ranking group are trained with weights that decrease with the ranks of each training sample.}
	\Description[The schematics showing the concept of proposed calibration method]{Several users have their own item rankings. The top-1 and top-2 items are surrounded by the dotted box, representing an extracted dataset for one ranking group. Similarly, the top-3 and top-4 items are surrounded by another dotted box. On the right of the dotted boxes, there are graphs showing the mapping functions from initial scores to calibrated scores. The two graphs for each two ranking groups represent the construction of distinct ranking models for each ranking model.}
	\label{fig:proposed_method}
\end{figure}

In this section, we propose an optimization method for calibration models to address the miscalibration in top-N items.
We assume that we have a recommender model that is already trained and a validation dataset that can be used for learning a calibration model.
We first rank items in the validation dataset using the recommender model.
Then, we extract the top-N items for each user and group them into $n_g$ groups by their ranks as shown in Fig. \ref{fig:proposed_method}.
The data in each group,  $\{y_k, s_k, r_k\}$, are triplets of a ground truth label, a recommender output score, and a rank.
Finally, we train calibration models for each ranking group using the data with discounting weights $w_k=(1/r_k)^\alpha$.

We expect that the separate calibration models address the rank-dependency of miscalibration, as shown in Fig. \ref{fig:reliability_diagram} (b).
Further, the rank-discounting weight is aimed at prioritizing the accuracy of calibration in higher-ranked items, which are often practically important in various recommendation scenarios.
Note that $n_g$ and $\alpha$ are hyperparameters of the proposed training method.
We set $n_g=N/5$ (i.e., 5 ranks in each group\footnote{In case N is not divisible by $n_g$, we can have different number of ranks in each group. For example, if $N=18$ and $n_g=4$, the number of ranks in each group can be 5, 4, 5, 4.}) and $\alpha=1$ as defaults.
We also investigate the sensitivity to these parameters in the experiment section.

\section{Experiments}
\subsection{Experimental Settings}
In this section, we concisely describe our experimental settings.

\textbf{Datasets}
We experiment on both explicit rating prediction and implicit preference prediction tasks.
For rating prediction, we used the popular MovieLens-1M (ML-1M) dataset~\cite{harper2015movielens}.
For preference prediction, we used the KuaiRec dataset~\cite{gao2022kuairec}.
The KuaiRec dataset is a fully-observed dataset, where 1,411 users have been exposed to all 3,327 items.
This dataset enables unbiased evaluation of the calibration quality, while most implicit feedback datasets are affected by exposure bias.\footnote{We chose the KuaiRec dataset instead of Yahoo!R3 and Coat datasets that are used in the previous work~\cite{kweon2022obtaining}. The KuaiRec dataset is much larger and the Yahoo!R3 and Coat datasets are insufficient for evaluating the calibration of top-N items.}

\textbf{Recommendation models}
To verify the versatility of the proposed method, we select popular recommendation models from diverse model classes.
For rating prediction, we employ the item-based KNN (itemKNN) algorithm~\cite{sarwar2001item} and the matrix factorization (MF) algorithm~\cite{koren2009matrix}.
They are representative methods from the neighborhood and latent factor approaches, respectively.
For preference prediction, we employ Bayesian personalized ranking (BPR)~\cite{rendle2009bpr}, neural collaborative filtering (NCF)~\cite{he2017neural}, and LightGCN~\cite{he2020lightgcn}.
They include both non-neural and neural recommendation models.
We used Cornac\footnote{https://github.com/PreferredAI/cornac}~\cite{salah2020cornac} for itemKNN, MF, and BPR, and Recommenders\footnote{https://github.com/recommenders-team/recommenders}~\cite{graham2019microsoft} for NCF and LightGCN, all with the default hyperparameters. 

\textbf{Calibration models}
We apply various non-parametric (histogram binning, isotonic regression) and parametric (Platt scaling, Beta calibration, Gaussian calibration, Gamma calibration) models.
We also evaluate as-is recommendation scores without any calibrations (refer to as Vanilla).
Since the parametric models are designed for calibrating scores to [0,1] probability, we evaluate only with non-parametric models for five-scale rating prediction task.

\textbf{Compared methods}
We compare our top-N focused optimization method (TNF) with two baselines.
The first one is to optimize calibrators for all user-item pairs (refer to as Original).
The second one is the variance-adjusting debiasing (VAD) method~\cite{fan2023calibration}, which is a recently proposed method to address overestimates in top-N items.

\textbf{Evaluation process}
We split a dataset into training, validation, and test datasets with the ratio of 60\%, 20\%, and 20\%, respectively.
We first train a recommendation model with the training dataset.
Then, we optimize calibration models using the validation dataset.
Finally, we evaluate ECE@N and RDECE@N in the test dataset.
We repeat the above process 10 times with different data splitting using varied random seeds and report the mean values of the evaluation metrics.
As for the number of recommendation, we set N=20 for most experiments.
We also evaluate with varied N.
To obtain more reliable estimates of ECE@N, following~\cite{roelofs2022mitigating}, we adjust the number of bins to be as large as possible under the monotonicity constraint.\footnote{We also evaluated with a fixed number of bins and obtained similar results.}

Note that the calibration does not affect the ranking of top-N items since we first extract top-N items before applying calibration.
Therefore, ranking-based metrics (e.g., AUC and NDCG) depend only on recommender models, independent of the choice of calibration methods and calibration models.
Due to this independence from calibration, we put the evaluation of the ranking metrics and RMSE in the appendix (see Table \ref{tab:eval_accuracy}).

\subsection{Results}

\begin{table*}[htbp]
	\small
	\caption{Calibration comparison in the ML-1M dataset with itemKNN and MF. The best results among compared methods (Original, VAD, and TNF) for each combination of recommenders and calibration models are highlighted in bold. }
	\label{tab:comparison_ML1M}
	\centering
	\scalebox{1.0}{
			\begin{tabular}{llccc|ccc}
					\hline
					& & \multicolumn{3}{c}{ECE@20} & \multicolumn{3}{c}{RDECE@20}\\
					\cmidrule(lr){3-5} \cmidrule(lr){6-8} 
					Recommender & Calibration model  & Original &  VAD & TNF  & Original &  VAD & TNF \\
					\hline
					itemKNN & Vanilla (as reference) &  
					0.075 $\pm$ 0.010 & 0.524 $\pm$ 0.013 & - & 
					0.069 $\pm$ 0.021 & 0.605 $\pm$ 0.037 & - \\ 
					 & Histogram binning & 
					0.194 $\pm$ 0.011 & 0.471 $\pm$ 0.010 & \textbf{0.045} $\pm$ 0.006 & 
					0.229 $\pm$ 0.045 & 0.537 $\pm$ 0.038 & \textbf{0.072} $\pm$ 0.022 \\ 
					 & Isotonic regression & 
					0.177 $\pm$ 0.011 & 0.462 $\pm$ 0.010 & \textbf{0.026} $\pm$ 0.015 & 
					0.197 $\pm$ 0.044 & 0.521 $\pm$ 0.038 & \textbf{0.070} $\pm$ 0.021 \\ 
					\hline
					MF & Vanilla (as reference) &  
					0.098 $\pm$ 0.014 & 0.327 $\pm$ 0.016 & - & 
					0.079 $\pm$ 0.020 & 0.332 $\pm$ 0.019 & - \\ 
					 & Histogram binning & 
					0.065 $\pm$ 0.013 & 0.351 $\pm$ 0.014 & \textbf{0.029} $\pm$ 0.010 & 
					0.089 $\pm$ 0.014 & 0.387 $\pm$ 0.014 & \textbf{0.027}  $\pm$ 0.008 \\ 
					 & Isotonic regression & 
					0.036 $\pm$ 0.012 & 0.325 $\pm$ 0.013 & \textbf{0.017} $\pm$ 0.008 & 
					0.047 $\pm$ 0.010 & 0.348 $\pm$ 0.013 & \textbf{0.027} $\pm$ 0.009 \\ 
					\hline
				\end{tabular}
		}
\end{table*}

\begin{table*}[htbp]
	\small
	%		\footnotesize
	\caption{Calibration comparison in the Kuairec dataset with NCF, lightGCN, and BPR. The best results among compared methods (Original, VAD, and TNF) for each combination of recommenders and calibration models are highlighted in bold.}
	\label{tab:comparison_KuaiRec}
	\centering
	\scalebox{1.0}{
		\begin{tabular}{llccc|ccc}
			\hline
			& & \multicolumn{3}{c}{ECE@20} & \multicolumn{3}{c}{RDECE@20}\\
			\cmidrule(lr){3-5} \cmidrule(lr){6-8} 
			Recommender & Calibration model  & Original &  VAD & TNF  & Original &  VAD & TNF \\
			\hline
			NCF & Vanilla (as reference)&  
			0.340 $\pm$ 0.013 & 0.171 $\pm$ 0.016 & - & 
			0.292 $\pm$ 0.018 & 0.115 $\pm$ 0.021 & - \\ 
			& Histogram binning & 
			0.260 $\pm$ 0.011 & 0.312 $\pm$ 0.011 & \textbf{0.020} $\pm$ 0.004 & 
			0.342 $\pm$ 0.014 & 0.394 $\pm$ 0.014 & \textbf{0.023}  $\pm$ 0.005 \\ 
			& Isotonic regression & 
			0.080 $\pm$ 0.017 & 0.152 $\pm$ 0.015 & \textbf{0.012} $\pm$ 0.006 & 
			0.116 $\pm$ 0.021 & 0.197 $\pm$ 0.021 & \textbf{0.023}  $\pm$ 0.005 \\ 
			& Platt scaling & 
			0.040 $\pm$ 0.008 & 0.096 $\pm$ 0.021 & \textbf{0.013} $\pm$ 0.003 & 
			0.070 $\pm$ 0.024 & 0.152 $\pm$ 0.028 & \textbf{0.023}  $\pm$ 0.005 \\ 
			& Beta calibration & 
			0.093 $\pm$ 0.018 & 0.147 $\pm$ 0.013 & \textbf{0.010}  $\pm$ 0.004 & 
			0.102 $\pm$ 0.017 & 0.183 $\pm$ 0.018 & \textbf{0.023} $\pm$ 0.005 \\ 
			& Gaussian calibration & 
			0.089 $\pm$ 0.024 & 0.139 $\pm$ 0.024 & \textbf{0.016} $\pm$ 0.005 & 
			0.093 $\pm$ 0.031 & 0.175 $\pm$ 0.032 & \textbf{0.023}  $\pm$ 0.005 \\ 
			& Gamma calibration & 
			0.087 $\pm$ 0.020 & 0.224 $\pm$ 0.038 & \textbf{0.013} $\pm$ 0.007 & 
			0.119 $\pm$ 0.029 & 0.279 $\pm$ 0.045 & \textbf{0.026} $\pm$ 0.006 \\ 
			\hline
			lightGCN & Vanilla (as reference)&  
			0.554 $\pm$ 0.008 & 0.446 $\pm$ 0.007  & - & 
			0.472 $\pm$ 0.007 & 0.363 $\pm$ 0.007  & - \\ 
			& Histogram binning & 
			0.257 $\pm$ 0.009 & 0.297 $\pm$ 0.009 & \textbf{0.019} $\pm$ 0.004 & 
			0.339 $\pm$ 0.008 & 0.380 $\pm$ 0.008 & \textbf{0.022} $\pm$ 0.003 \\ 
			& Isotonic regression & 
			0.070 $\pm$ 0.009 & 0.128 $\pm$ 0.008 & \textbf{0.014} $\pm$ 0.004 & 
			0.078 $\pm$ 0.010 & 0.152 $\pm$ 0.009 & \textbf{0.022} $\pm$ 0.004 \\ 
			& Platt scaling & 
			0.132 $\pm$ 0.009 & 0.184 $\pm$ 0.008 & \textbf{0.018} $\pm$ 0.006 & 
			0.164 $\pm$ 0.009 & 0.227 $\pm$ 0.008 & \textbf{0.022} $\pm$ 0.004 \\ 
			& Beta calibration & 
			0.097 $\pm$ 0.009 & 0.145 $\pm$ 0.008 & \textbf{0.015} $\pm$ 0.007 & 
			0.092 $\pm$ 0.010 & 0.165 $\pm$ 0.009 & \textbf{0.022} $\pm$ 0.004 \\ 
			& Gaussian calibration & 
			0.096 $\pm$ 0.011 & 0.152 $\pm$ 0.012 & \textbf{0.016} $\pm$ 0.005 & 
			0.112 $\pm$ 0.013 & 0.182 $\pm$ 0.013 & \textbf{0.022} $\pm$ 0.004 \\ 
			& Gamma calibration & 
			0.131 $\pm$ 0.011 & 0.185 $\pm$ 0.011 & \textbf{0.015} $\pm$ 0.005 & 
			0.164 $\pm$ 0.012 & 0.228 $\pm$ 0.012 & \textbf{0.022} $\pm$ 0.004 \\
			\hline
			BPR & Vanilla (as reference)&  
			0.446 $\pm$ 0.005 & 0.429 $\pm$ 0.005  & - & 
			0.361 $\pm$ 0.004 & 0.344 $\pm$ 0.004  & - \\ 
			& Histogram binning & 
			0.291 $\pm$ 0.006 & 0.306 $\pm$ 0.006 & \textbf{0.018} $\pm$ 0.004 & 
			0.383 $\pm$ 0.005 & 0.399 $\pm$ 0.004 & \textbf{0.020} $\pm$ 0.003 \\ 
			& Isotonic regression & 
			0.034 $\pm$ 0.006 & 0.046 $\pm$ 0.006 & \textbf{0.011} $\pm$ 0.005 & 
			0.064 $\pm$ 0.004 & 0.077 $\pm$ 0.004 & \textbf{0.019} $\pm$ 0.004 \\ 
			& Platt scaling & 
			0.143 $\pm$ 0.006 & 0.153 $\pm$ 0.006 & \textbf{0.016} $\pm$ 0.004 & 
			0.186 $\pm$ 0.005 & 0.197 $\pm$ 0.005 & \textbf{0.020} $\pm$ 0.003 \\ 
			& Beta calibration & 
			0.072 $\pm$ 0.007 & 0.076 $\pm$ 0.007 & \textbf{0.013} $\pm$ 0.004 & 
			0.039 $\pm$ 0.006 & 0.051 $\pm$ 0.007 & \textbf{0.020} $\pm$ 0.003 \\ 
			& Gaussian calibration & 
			0.069 $\pm$ 0.011 & 0.078 $\pm$ 0.012 & \textbf{0.016} $\pm$ 0.004 & 
			0.079 $\pm$ 0.015 & 0.093 $\pm$ 0.015 & \textbf{0.020} $\pm$ 0.003 \\ 
			& Gamma calibration & 
			0.086 $\pm$ 0.014 & 0.101 $\pm$ 0.012 & \textbf{0.016} $\pm$ 0.003 & 
			0.119 $\pm$ 0.015 & 0.137 $\pm$ 0.013 & \textbf{0.020} $\pm$ 0.003 \\ 
			\hline
		\end{tabular}
	}
\end{table*}

In this section, we explain and discuss the experiment results.

\textbf{Performance comparison}
Tables \ref{tab:comparison_ML1M}  and \ref{tab:comparison_KuaiRec} show the comparisons of top-N calibration.
For each calibration model, TNF achieves lower calibration errors compared with Original and VAD. 
In the ML-1M dataset with itemKNN (Table \ref{tab:comparison_ML1M}), the results of histogram binning and Isotonic regression are worse than that of the Vanilla, meaning that the application of calibration models optimized in the original method (i.e., trained on all items) degrades the calibration performance.\footnote{Note that such miscalibration happens only for top-N items and the application of calibrators to itemKNN improves the conventional ECE. Please refer to Table \ref{tab:conventional_ECE} in the appendix. Also, the original calibrators are better than Vanilla in ML-1M with MF as shown in Table \ref{tab:comparison_ML1M}.}
On the contrary, TNF outperforms Vanilla in ECE@20 and comparable in RDECE@20.
We find that VAD increases calibration errors in almost all cases.
VAD aims to compensate for the overestimates in top-N calibration; however, underestimates are also common in top-N, as shown in Fig. \ref*{fig:reliability_diagram}.
In such cases, VAD might have an adverse effect.

\begin{figure}[htbp]
	\begin{center}
		\subfigure[Isotonic regression with itemKNN]{\includegraphics[width=0.23\textwidth]{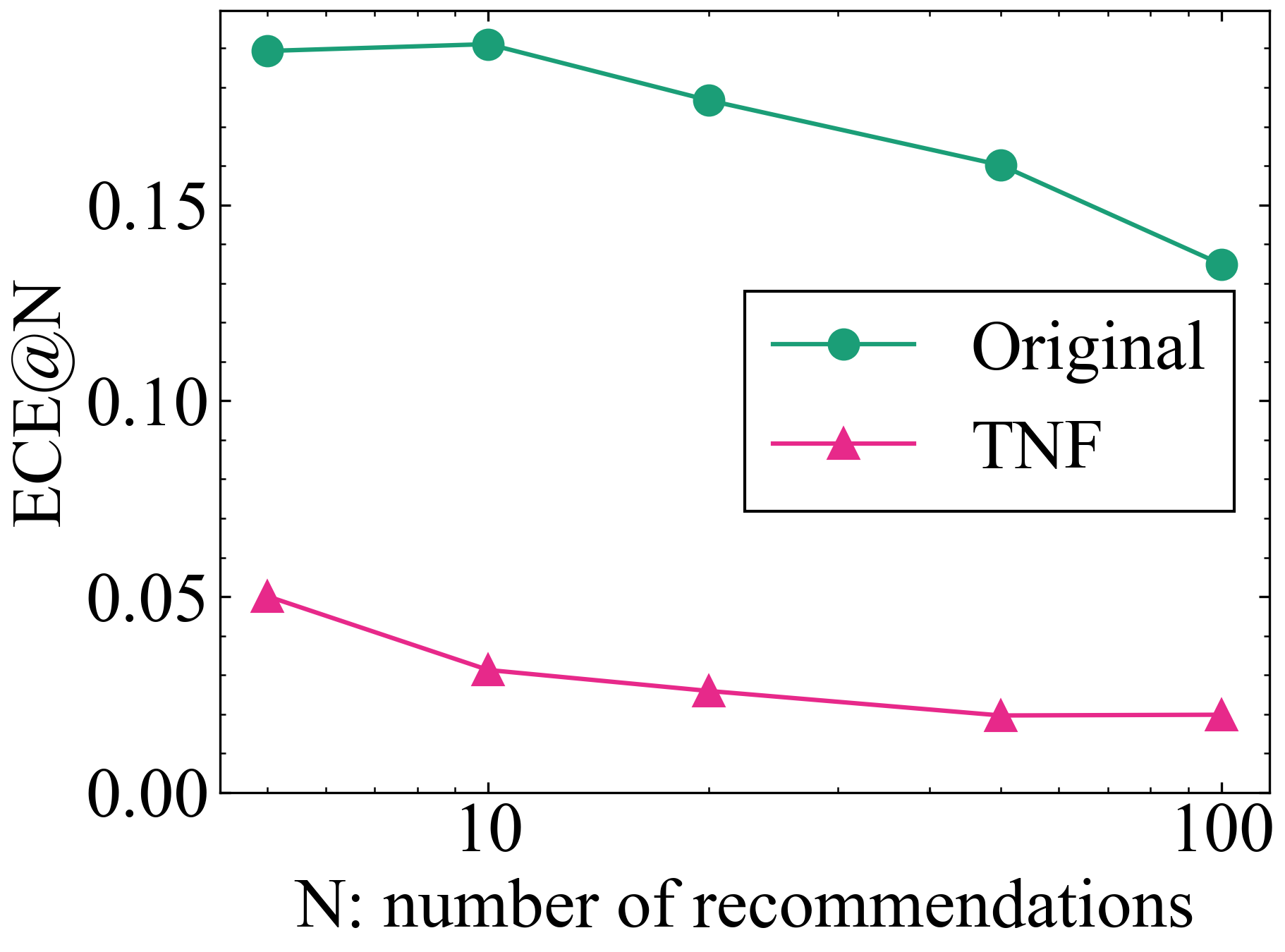}}
		\Description[Line showing the expected calibration errors at the top-N items]{Line graph showing the expected calibration errors at the top-N items from 0.0 to 0.18 on the Y axis against the number of recommendations N from 5 to 100 on the X axis. The plots for proposed top-N focused calibrator optimization method have ECE lower than the plots for original calibrator optimization method using all items' data.}
		\subfigure[Isotonic regression with MF]{\includegraphics[width=0.23\textwidth]{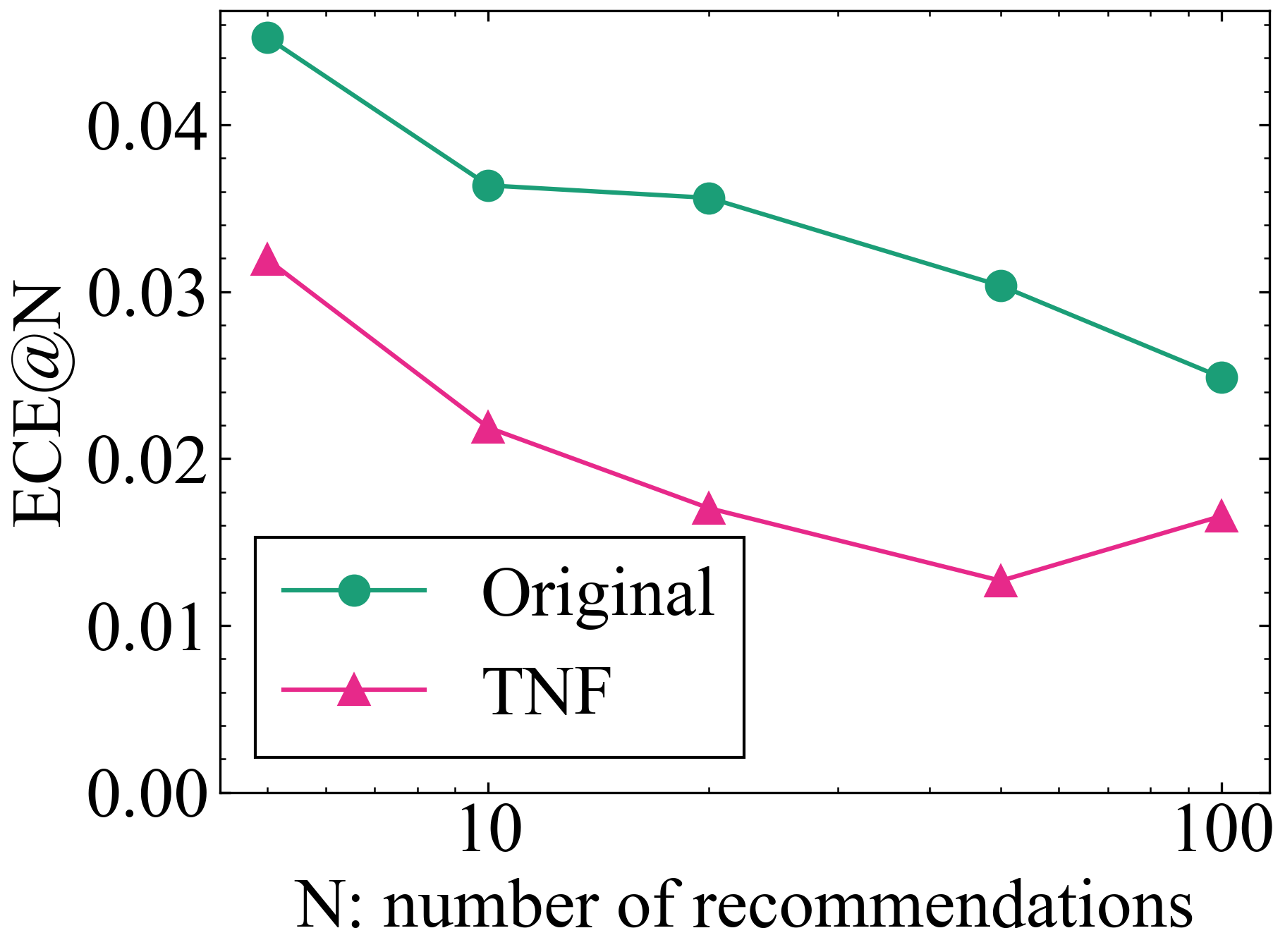}}
		\Description[Line showing the expected calibration errors at the top-N items]{Line graph showing the expected calibration errors at the top-N items from 0.0 to 0.05 on the Y axis against the number of recommendations N from 5 to 100 on the X axis. The plots for proposed top-N focused calibrator optimization method have ECE lower than the plots for original calibrator optimization method using all items' data.}
		\subfigure[Beta calibration with NCF]{\includegraphics[width=0.23\textwidth]{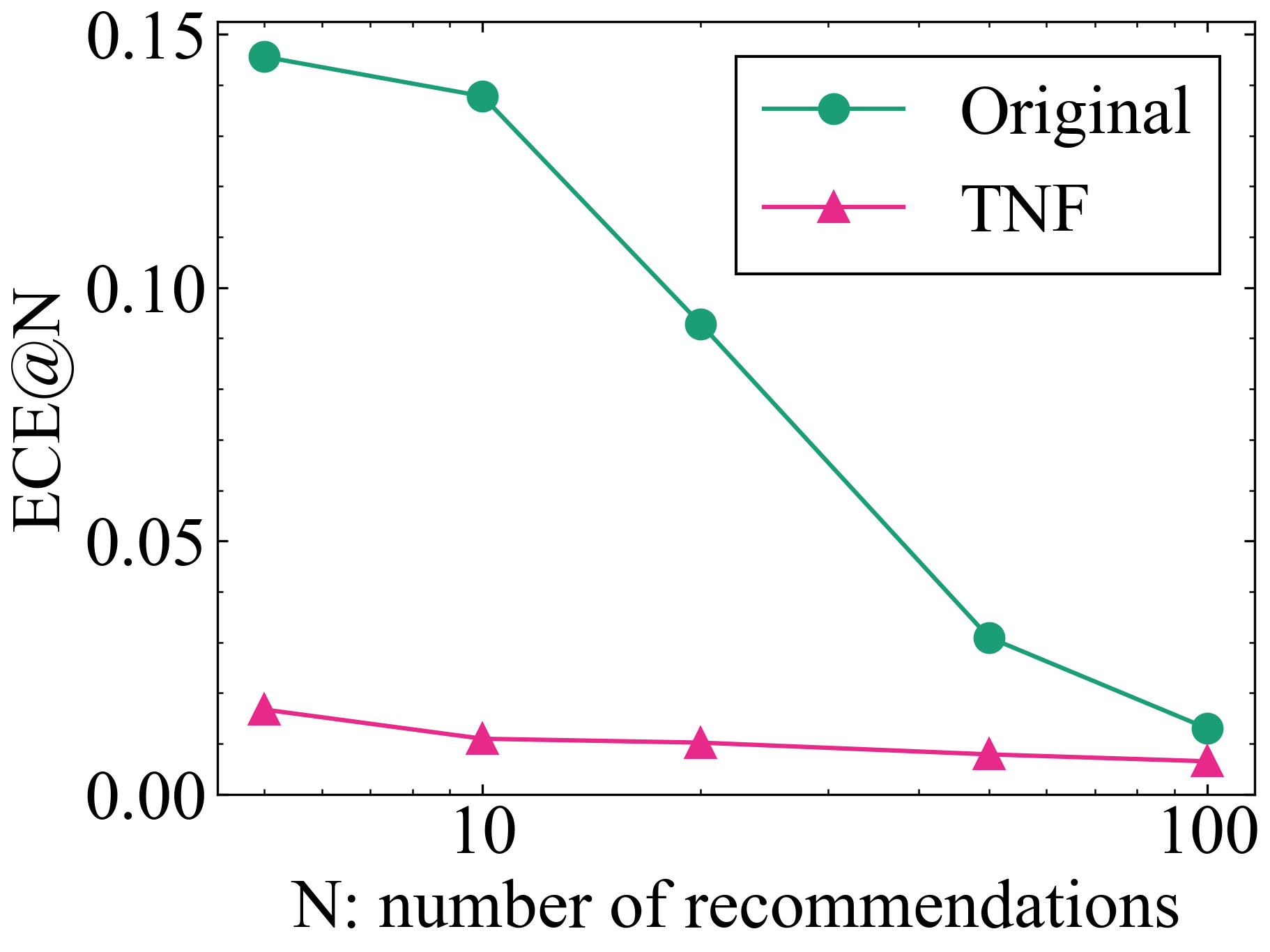}}
		\Description[Line showing the expected calibration errors at the top-N items]{Line graph showing the expected calibration errors at the top-N items from 0.0 to 0.15 on the Y axis against the number of recommendations N from 5 to 100 on the X axis. The plots for proposed top-N focused calibrator optimization method have ECE lower than the plots for original calibrator optimization method using all items' data.}
		\subfigure[Gamma calibration with LightGCN]{\includegraphics[width=0.23\textwidth]{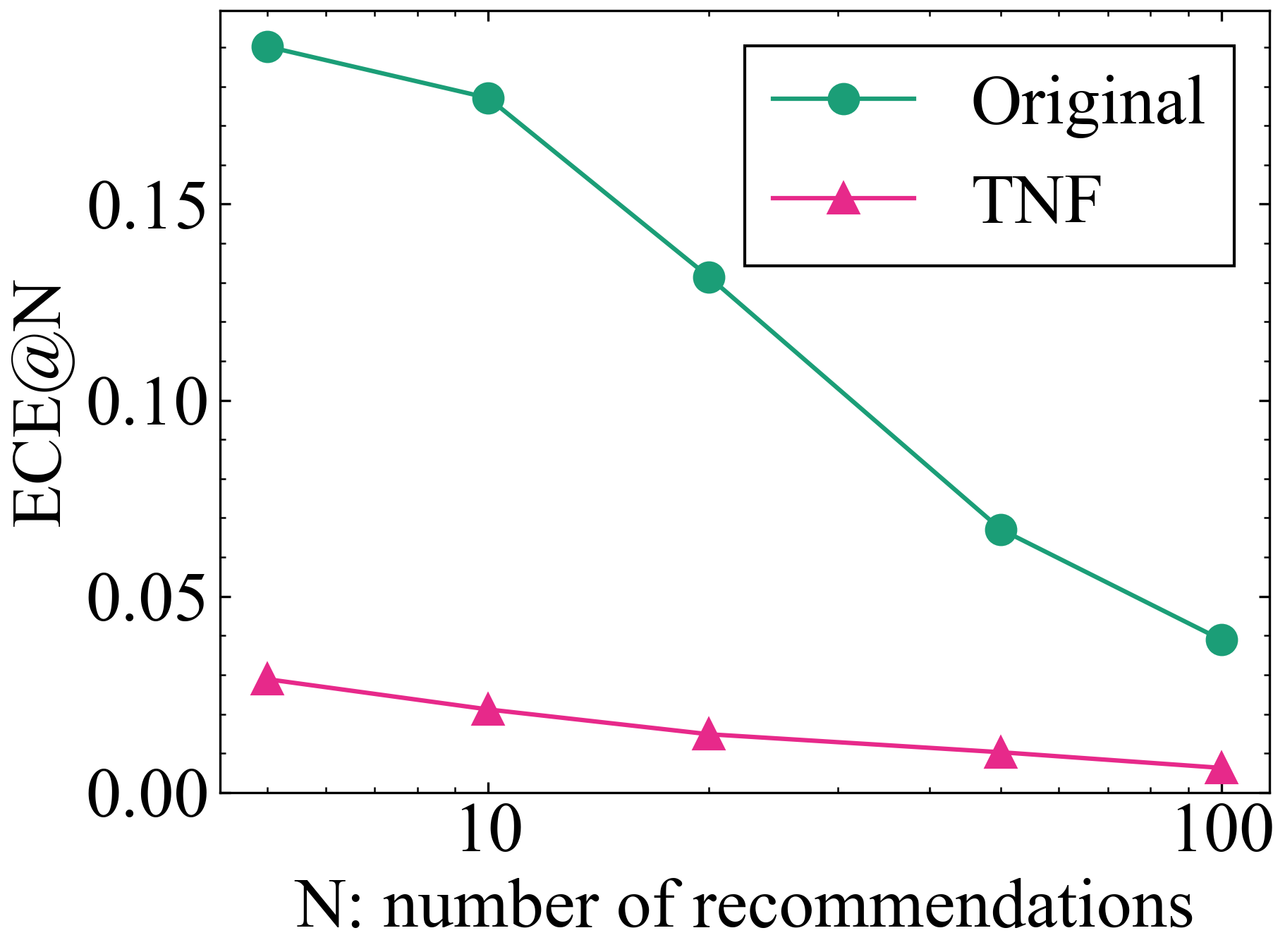}}
		\Description[Line showing the expected calibration errors at the top-N items]{Line graph showing the expected calibration errors at the top-N items from 0.0 to 0.18on the Y axis against the number of recommendations N from 5 to 100 on the X axis. The plots for proposed top-N focused calibrator optimization method have ECE lower than the plots for original calibrator optimization method using all items' data.}
%		\subfigure[Isotonic regression with BPR]{\includegraphics[width=0.23\textwidth]{figures/2024-05-12_dep_nrec_KuaiRec_BPR_isotonic.png}}
%		\Description[Line showing the expected calibration errors at the top-N items]{Line graph showing the expected calibration errors at the top-N items from 0.0 to 0.08 on the Y axis against the number of recommendations N from 5 to 100 on the X axis. The plots for proposed top-N focused calibrator optimization method have ECE lower than the plots for original calibrator optimization method using all items' data.}
		\caption{ECE@N for varied number of recommendations.}
		\label{fig:dep_nrec}
	\end{center}
\end{figure}

\textbf{Dependence on the number of recommendation}
Next we investigate the trends of ECE@N for a varying number of recommendations.
As shown in Fig. \ref{fig:dep_nrec}, ECE@N increases as the number of recommendations decreases.
TNF effectively suppresses the miscalibration compared to the original calibrators.

\begin{figure}[b]
	\begin{center}
		\subfigure[ECE@20 in ML-1M]{\includegraphics[width=0.22\textwidth]{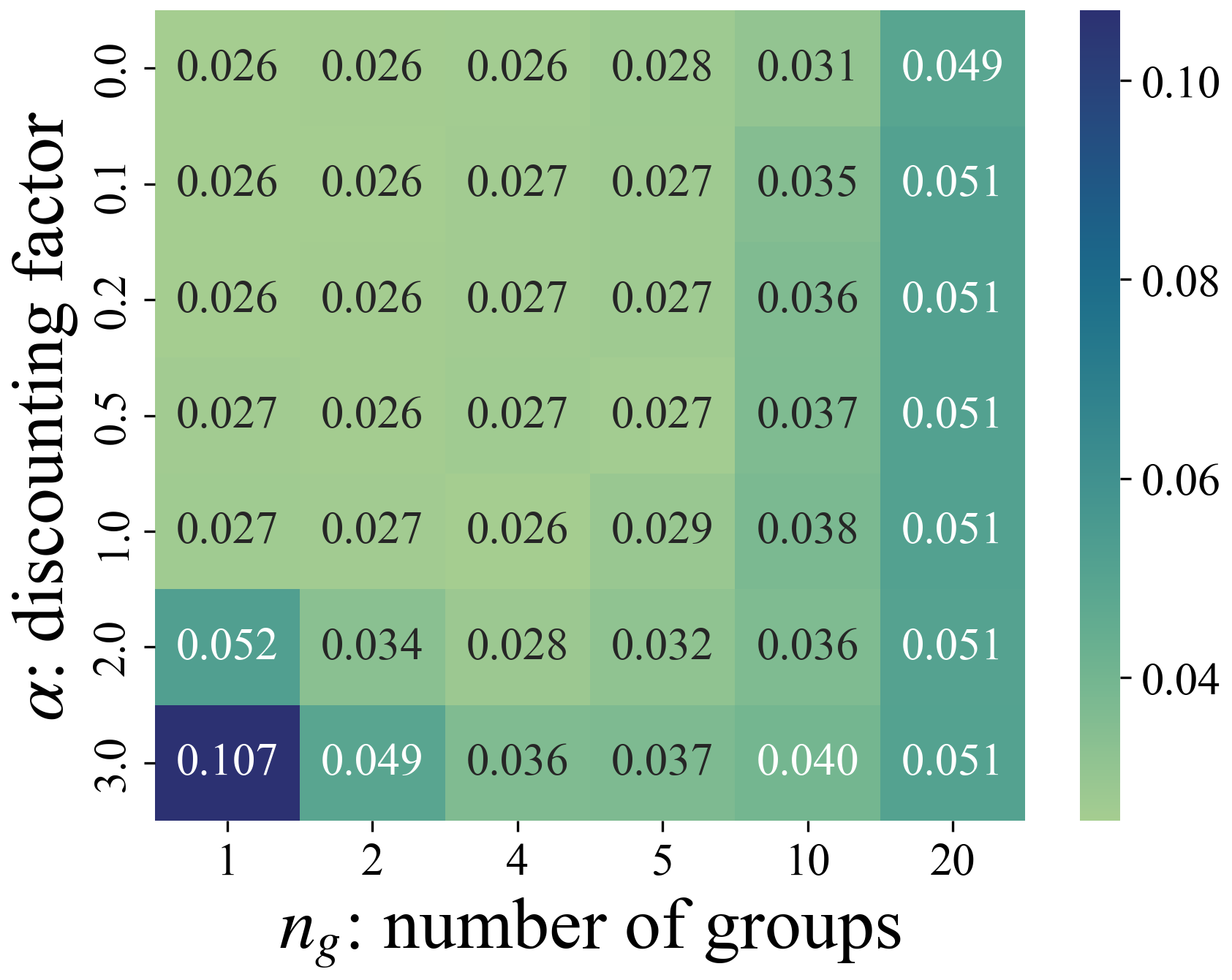}}
		\Description[Heat map showing the expected calibration error at top-N items.]{Heat map showing the expected calibration error at top-N items with color scale from 0.04 to 0.10 and the discounting factor from 3.0 to 0.0 on the Y axis against the number of groups from 1 to 20 on the X axis.}
		\subfigure[RDECE@20 in ML-1M]{\includegraphics[width=0.22\textwidth]{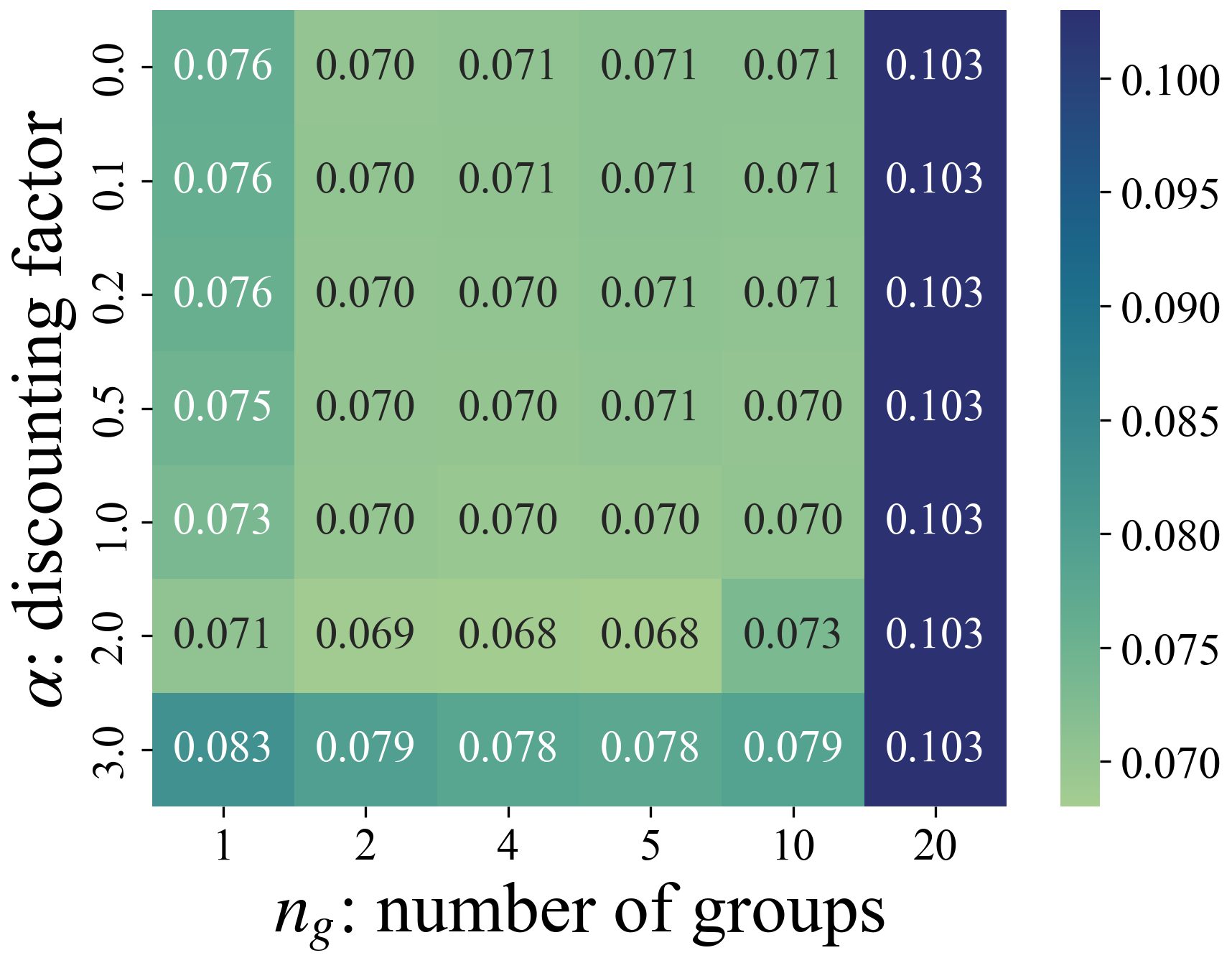}}
		\Description[Heat map showing the rank-discounted expected calibration error at top-N items.]{Heat map showing the rank-discounted expected calibration error at top-N items with color scale from 0.07 to 0.10 and the discounting factor from 3.0 to 0.0 on the Y axis against the number of groups from 1 to 20 on the X axis.}
		\subfigure[ECE@20 in KuaiRec]{\includegraphics[width=0.22\textwidth]{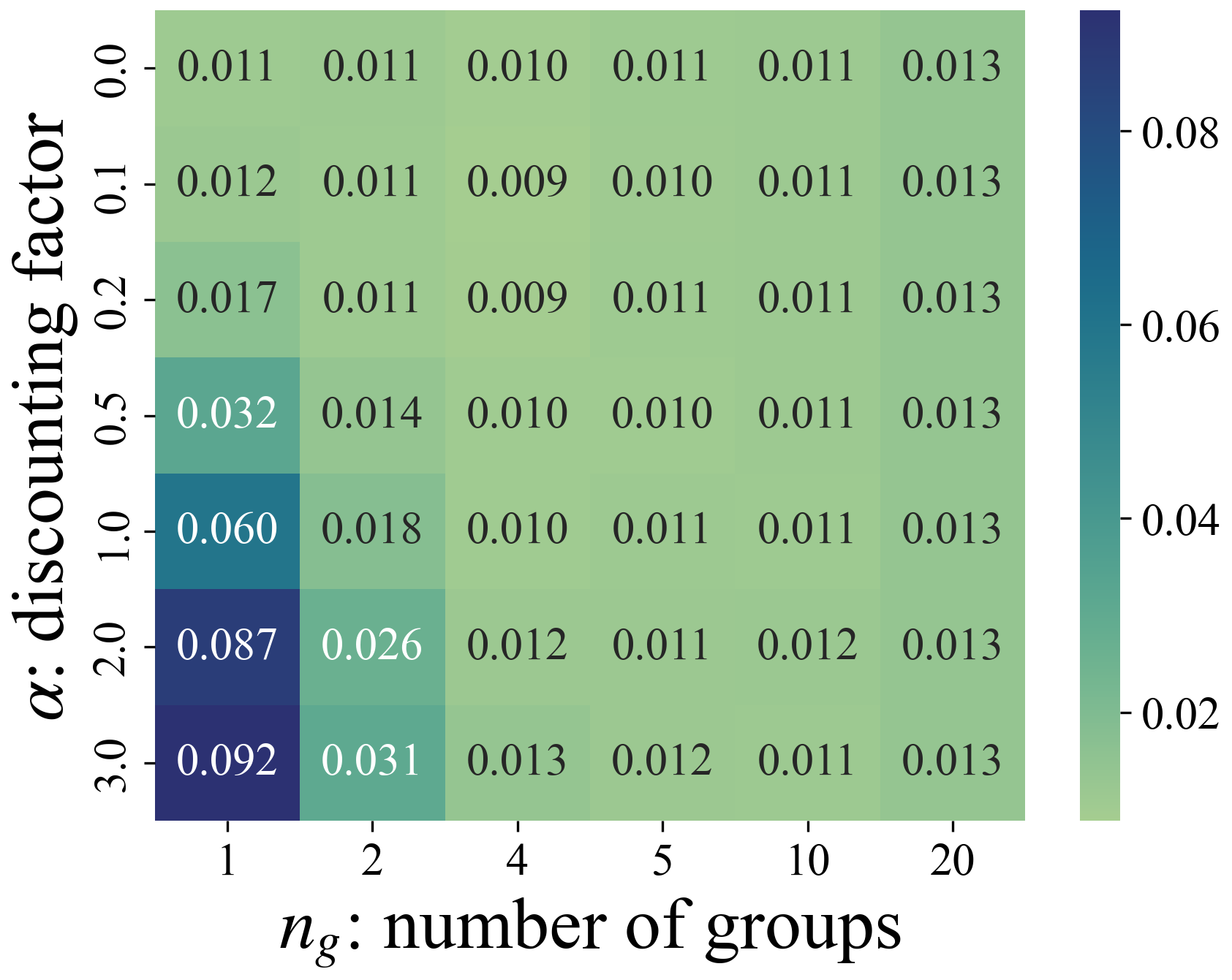}}
		\Description[Heat map showing the expected calibration error at top-N items.]{Heat map showing the expected calibration error at top-N items with color scale from 0.02 to 0.08 and the discounting factor from 3.0 to 0.0 on the Y axis against the number of groups from 1 to 20 on the X axis.}
		\subfigure[RDECE@20 in KuaiRec]{\includegraphics[width=0.22\textwidth]{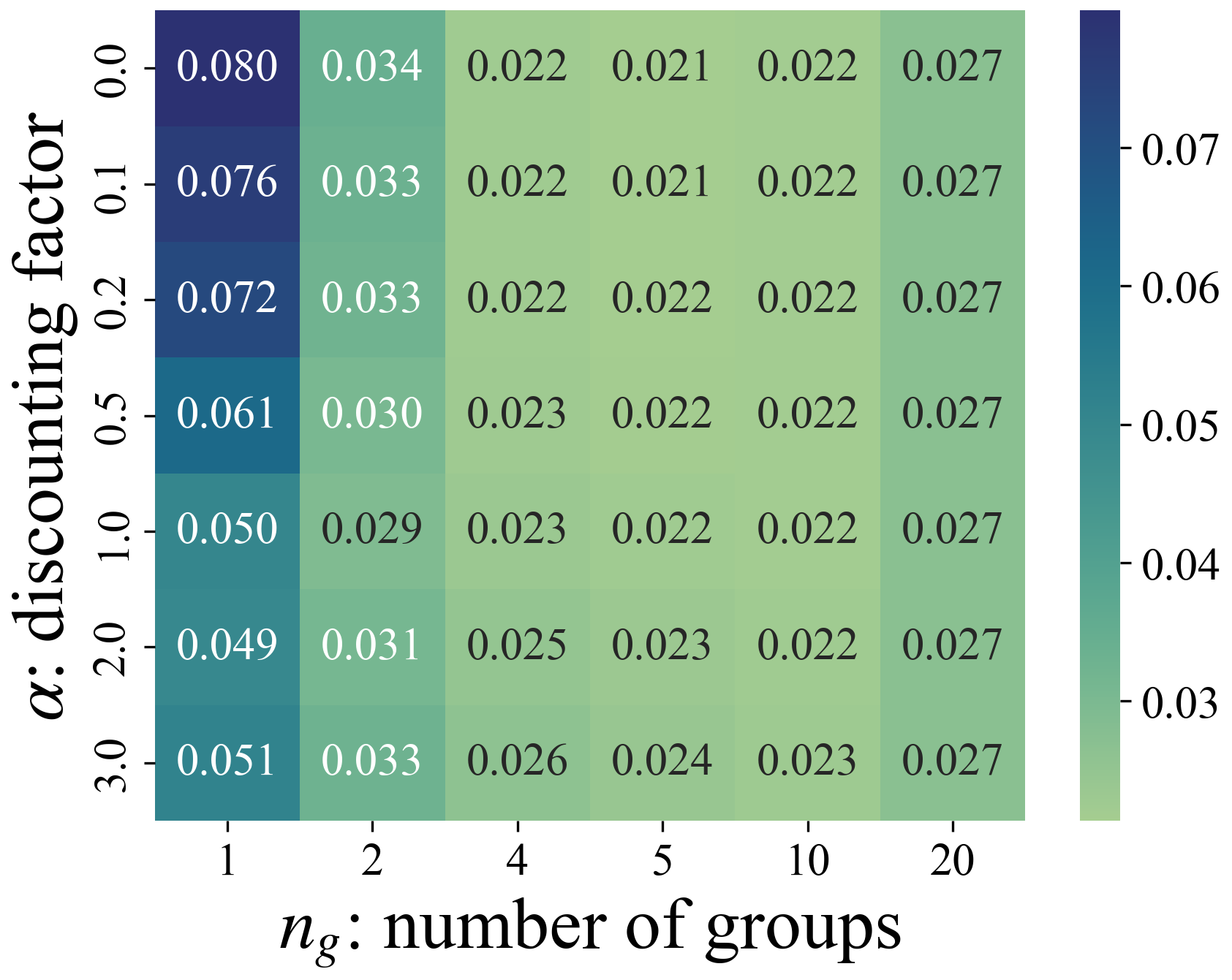}}
		\Description[Heat map showing the rank-discounted expected calibration error at top-N items.]{Heat map showing the rank-discounted expected calibration error at top-N items with color scale from 0.03 to 0.07 and the discounting factor from 3.0 to 0.0 on the Y axis against the number of groups from 1 to 20 on the X axis.}
		\caption{The sensitivities to the discounting factor $\alpha$ and the number of groups $n_g$. Recommender models are  ItemKNN and NCF and calibration models are isotonic regression and Beta calibration in ML-1M and KuaiRec, respectively.}
		\label{fig:sens_hp}
	\end{center}
\end{figure}

\textbf{Sensitivity to hyperparameters}
The proposed TNF has two hyperparameters: the discounting factor $\alpha$ to adjust the rank-based weights and the number of ranking groups $n_g$ to prepare separate calibration models.
Fig. \ref{fig:sens_hp} shows how ECE@N and RDECE@N are sensitive to those parameters\footnote{Additional results are in Fig.  \ref{fig:sens_hp_additional} in the appendix.}.
Note that we set  $n_g=N/5=4$ and $\alpha=1.0$ as default parameters in other experiments.
The optimal value for $\alpha$ is around 0.2 for ECE@N, and around 1.0 for RDECE@N.
This is reasonable considering that ECE@N treats all top-N items equally, while RDECE@N prioritizes higher-ranked items.
The optimal value for $n_g$ is around 5, showing that building calibration models for each ranking group is better than using one common model ($n_g=1$) or models for each rank ($n_g=20$).
On the other hand, unless extreme values are used (e.g., $\alpha=3.0$ or $n_g=1, 20$), TNF is not highly sensitive to these parameters and consistently improves upon the original method.

\section{Conclusions}
In this paper, we proposed evaluation metrics and a calibrator optimization method to address the miscalibration in top-N items.
We demonstrated the effectiveness of the proposed optimization method for both rating prediction and preference prediction tasks with various recommender models and calibration models.

%\clearpage
\bibliographystyle{ACM-Reference-Format}
\bibliography{reference_recsys2024}

\clearpage
\appendix
\section{Implementation Details}
We used the scikit-learn library\footnote{https://scikit-learn.org/} for Platt scaling (i.e., logistic regression) and Isotonic regression.
We used the betacal library\footnote{https://pypi.org/project/betacal/} for Beta calibration.
Following the official code,\footnote{https://github.com/WonbinKweon/CalibratedRankingModels\_AAAI2022} we implemented Gaussian calibration and Gamma calibration using the clogistic library.\footnote{https://github.com/guillermo-navas-palencia/clogistic}
Since all the above libraries accept sample weights, we directly input the rank-dependent weights for TNF optimization.
As for histogram binning, we first sort samples by scores and split them into 15 bins with an equal number of samples.
Then we take averages of ground truth labels for each bin.
In case of TNF, we take the weighted average with the rank-dependent weights.
In Vanilla calibration, we apply the sigmoid function for LightGCN and BPR since they outputs recommendation scores not bounded in [0, 1].
Conversely, the Cornac implementation of NCF outputs [0, 1] bounded scores, while Gaussian and Gamma calibrations expect unbounded recommendations scores.
Hence we conduct logit transformation to NCF scores before applying Gaussian or Gamma calibrations.
In Table \ref{tab:conventional_ECE}, we summarize the calibration performance in terms of conventional ECE (i.e., based on all items, not only top-N items) of each calibration model with the original method (i.e., training calibrators for all items).

\begin{table}[htbp]
	\caption{The results of conventional ECE.}
	%	\small
	\footnotesize
	\label{tab:conventional_ECE}
	\centering
	\scalebox{1.0}{
		\begin{tabular}{lcc|ccc}
			\hline
			& \multicolumn{2}{c}{ML-1M} & \multicolumn{3}{c}{KuaiRec}\\
			\cmidrule(lr){2-3} \cmidrule(lr){4-6} 
			Calibration model  & ItemKNN &  MF  & NCF &  LightGCN & BPR \\
			\hline
			Vanilla &  
			0.1427  & 0.0218  & 
			0.0779  & 0.5429  & 0.8511  \\ 
			Histogram binning & 
			0.0090  & 0.0079 & 
			0.0006 & 0.0006  & 0.0005  \\ 
			Isotonic regression & 
			0.0059  & 0.0075  & 
			0.0006  & 0.0005 & 0.0005 \\ 
			Platt scaling & 
			- & - &  
			0.0048 & 0.0037  & 0.0054  \\ 
			Beta calibration & 
			- & - & 
			0.0014  & 0.0017  & 0.0017  \\ 
			Gaussian calibration & 
			- & - & 
			0.0015 & 0.0025  & 0.0020  \\ 
			Gamma calibration & 
			- & - & 
			0.0011  & 0.0032  & 0.0034  \\
			\hline
		\end{tabular}
	}
\end{table}

\section{Additional experiment results}
\begin{figure}[htbp]
	\begin{center}
		\subfigure[Isotonic regression with BPR]{\includegraphics[width=0.25\textwidth]{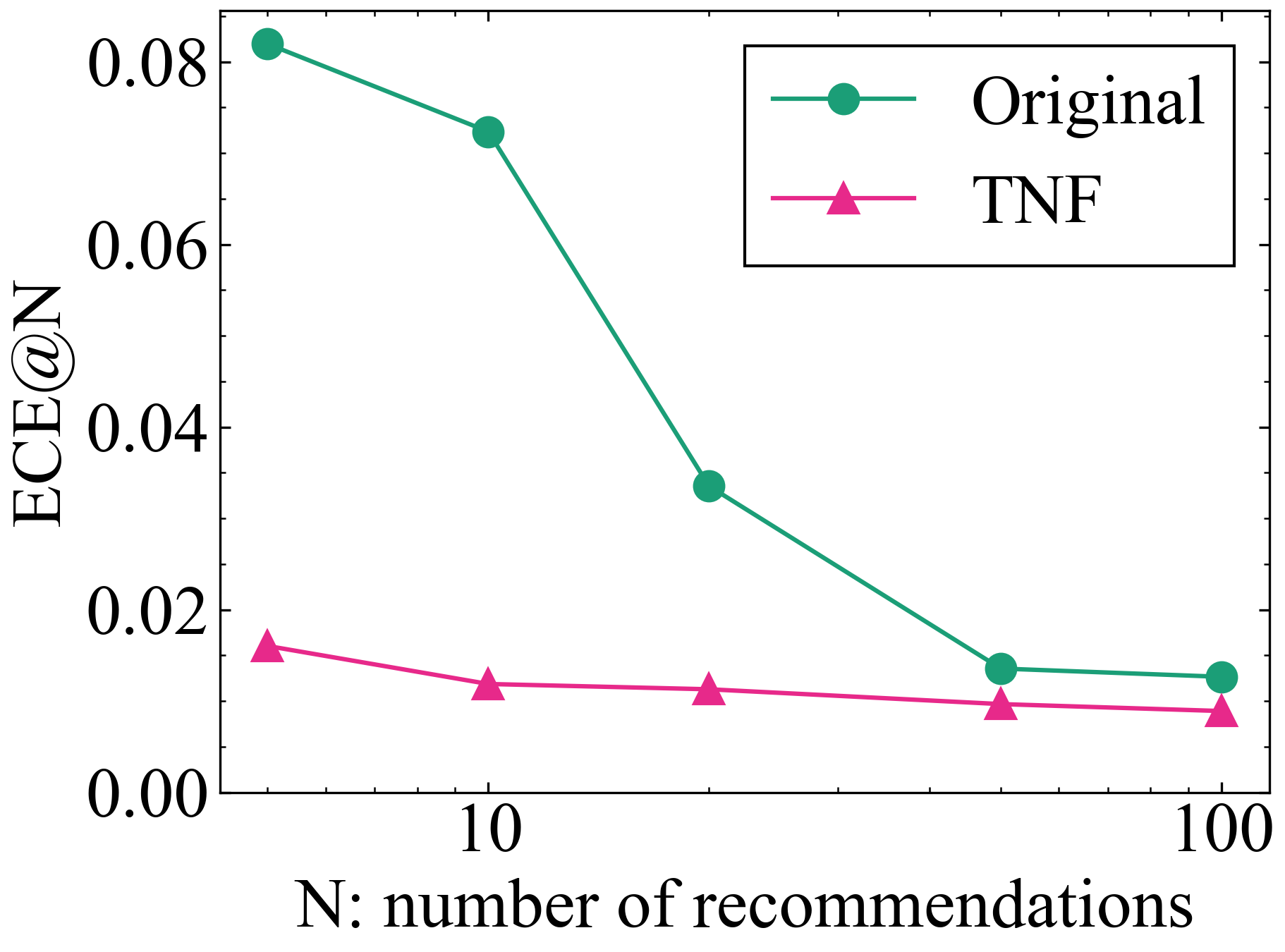}}
		\Description[Line showing the expected calibration errors at the top-N items]{Line graph showing the expected calibration errors at the top-N items from 0.0 to 0.08 on the Y axis against the number of recommendations N from 5 to 100 on the X axis. The plots for proposed top-N focused calibrator optimization method have ECE lower than the plots for original calibrator optimization method using all items' data.}
		\caption{ECE@N for varied N.}
		\label{fig:dep_nrec_additional}
	\end{center}
\end{figure}

\begin{figure}[htbp]
	\begin{center}
		\subfigure[ECE@20 for MF]{\includegraphics[width=0.22\textwidth]{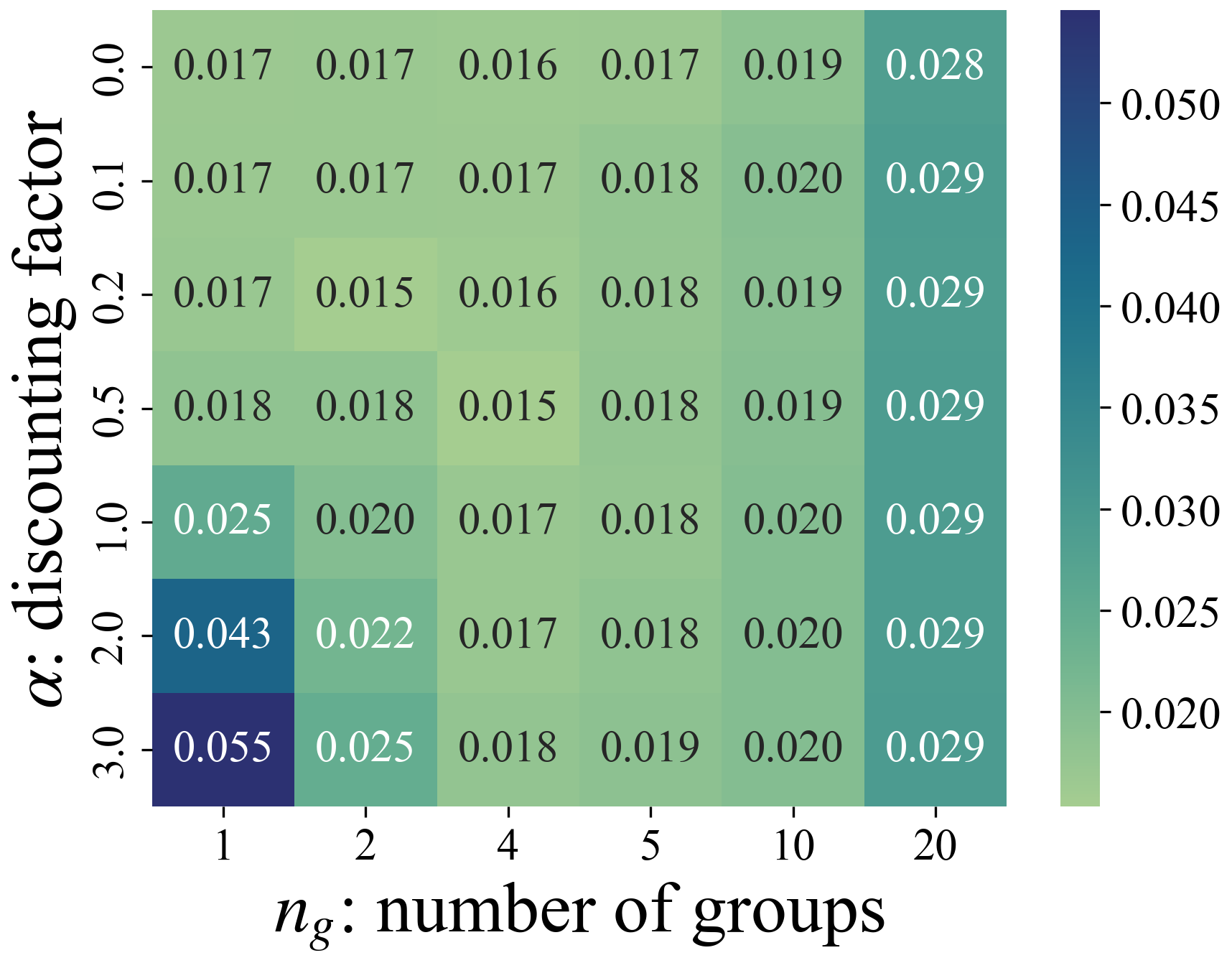}}
		\Description[Heat map showing the expected calibration error at top-N items.]{Heat map showing the expected calibration error at top-N items with color scale from 0.02 to 0.05 and the discounting factor from 3.0 to 0.0 on the Y axis against the number of groups from 1 to 20 on the X axis.}
		\subfigure[RDECE@20 for MF]{\includegraphics[width=0.22\textwidth]{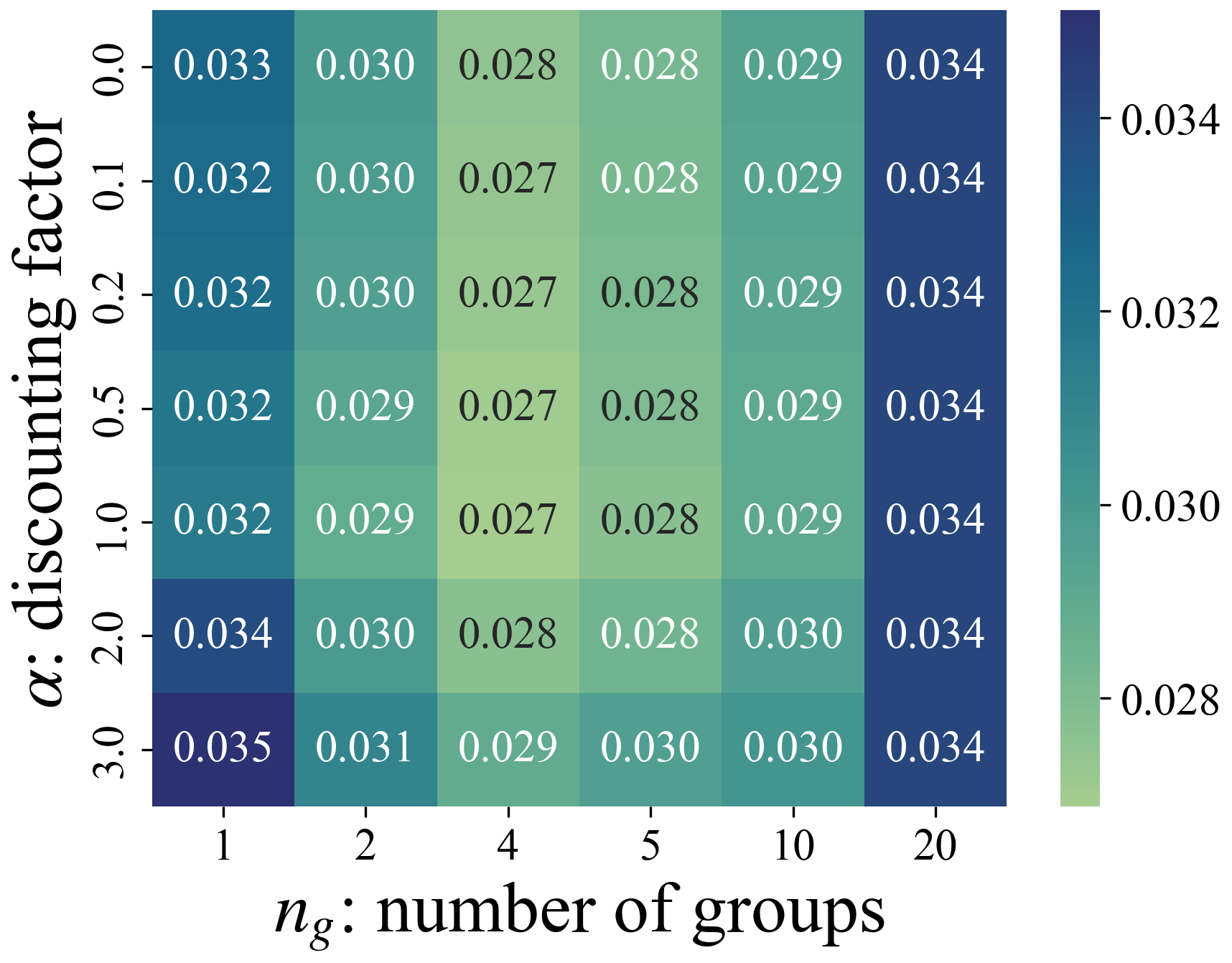}}
		\Description[Heat map showing the rank-discounted expected calibration error at top-N items.]{Heat map showing the rank-discounted expected calibration error at top-N items with color scale from 0.028 to 0.034 and the discounting factor from 3.0 to 0.0 on the Y axis against the number of groups from 1 to 20 on the X axis.}
		\subfigure[ECE@20 for LightGCN]{\includegraphics[width=0.22\textwidth]{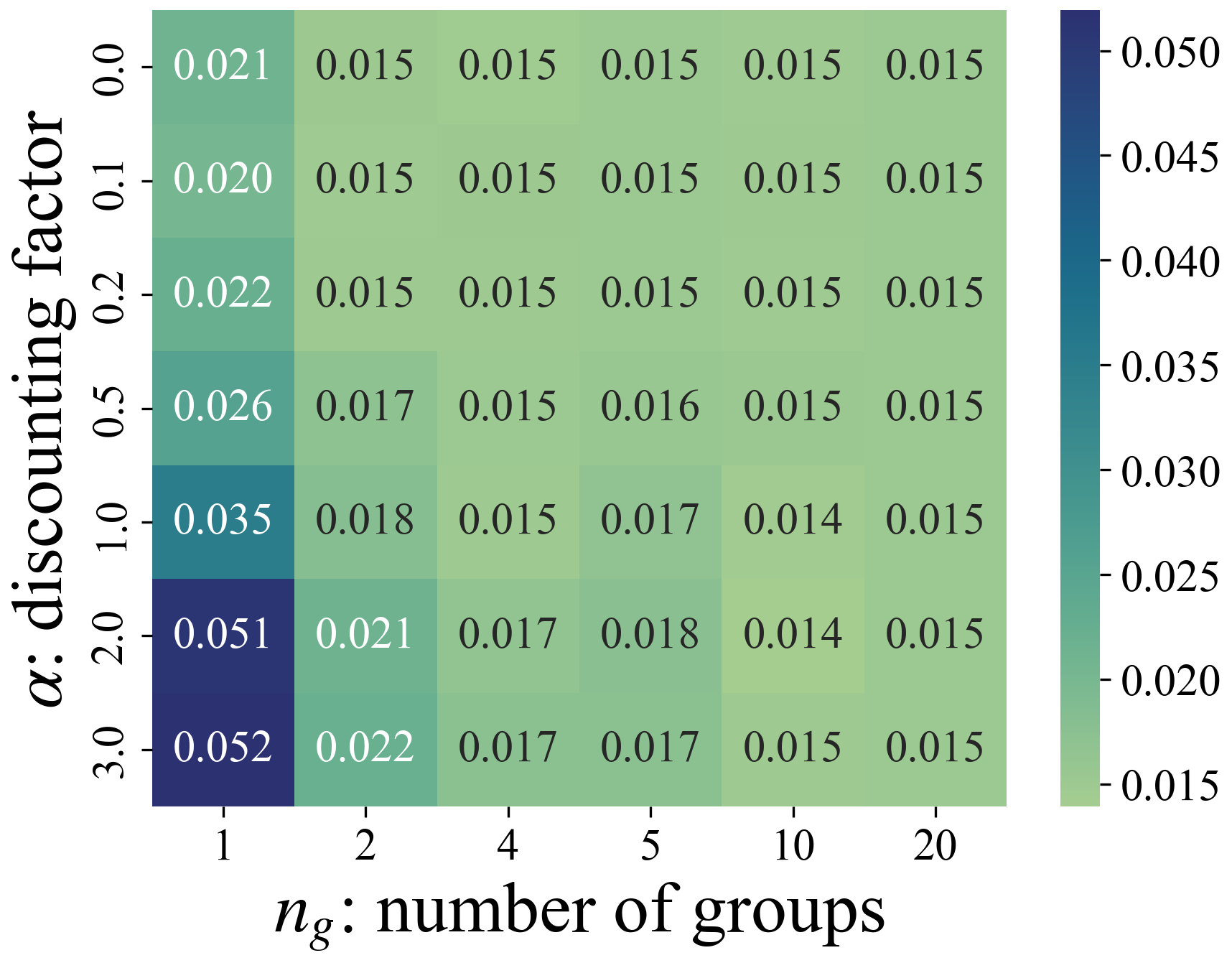}}
		\Description[Heat map showing the expected calibration error at top-N items.]{Heat map showing the expected calibration error at top-N items with color scale from 0.015 to 0.050 and the discounting factor from 3.0 to 0.0 on the Y axis against the number of groups from 1 to 20 on the X axis.}
		\subfigure[RDECE@20 for LightGCN]{\includegraphics[width=0.22\textwidth]{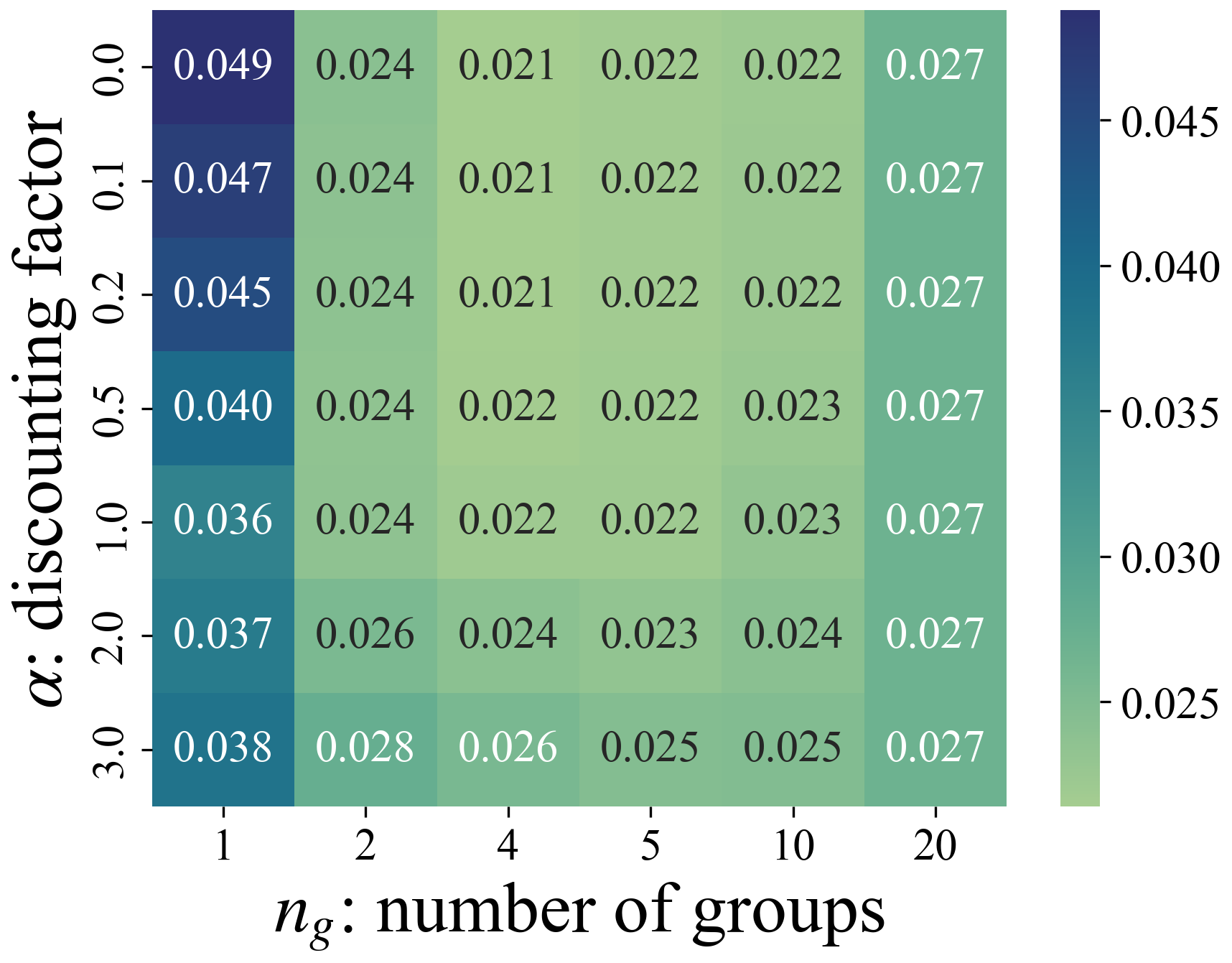}}
		\Description[Heat map showing the rank-discounted expected calibration error at top-N items.]{Heat map showing the rank-discounted expected calibration error at top-N items with color scale from 0.025 to 0.045 and the discounting factor from 3.0 to 0.0 on the Y axis against the number of groups from 1 to 20 on the X axis.}
		\subfigure[ECE@20 for BPR]{\includegraphics[width=0.22\textwidth]{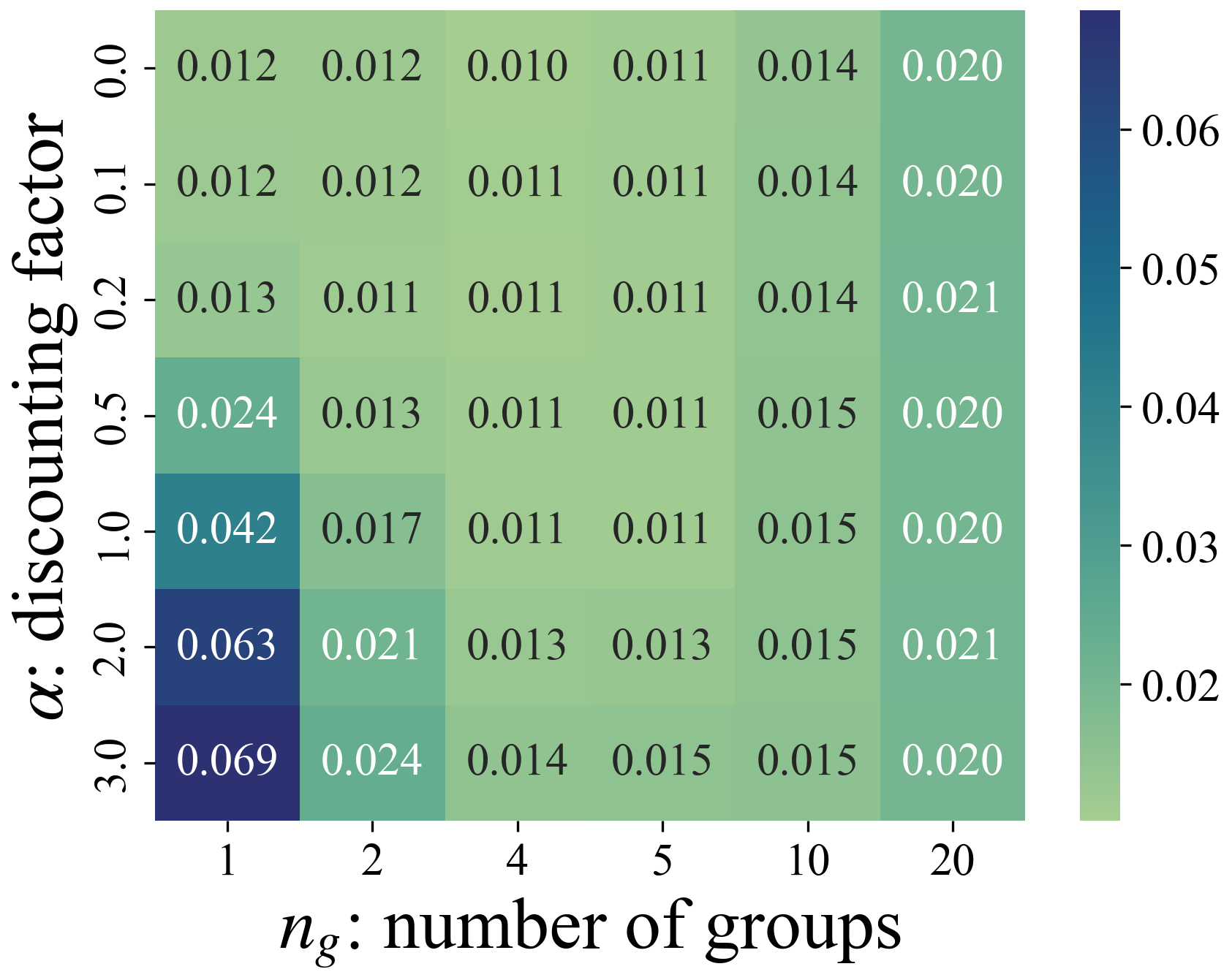}}
		\Description[Heat map showing the expected calibration error at top-N items.]{Heat map showing the expected calibration error at top-N items with color scale from 0.02 to 0.06 and the discounting factor from 3.0 to 0.0 on the Y axis against the number of groups from 1 to 20 on the X axis.}
		\subfigure[RDECE@20 for BPR]{\includegraphics[width=0.22\textwidth]{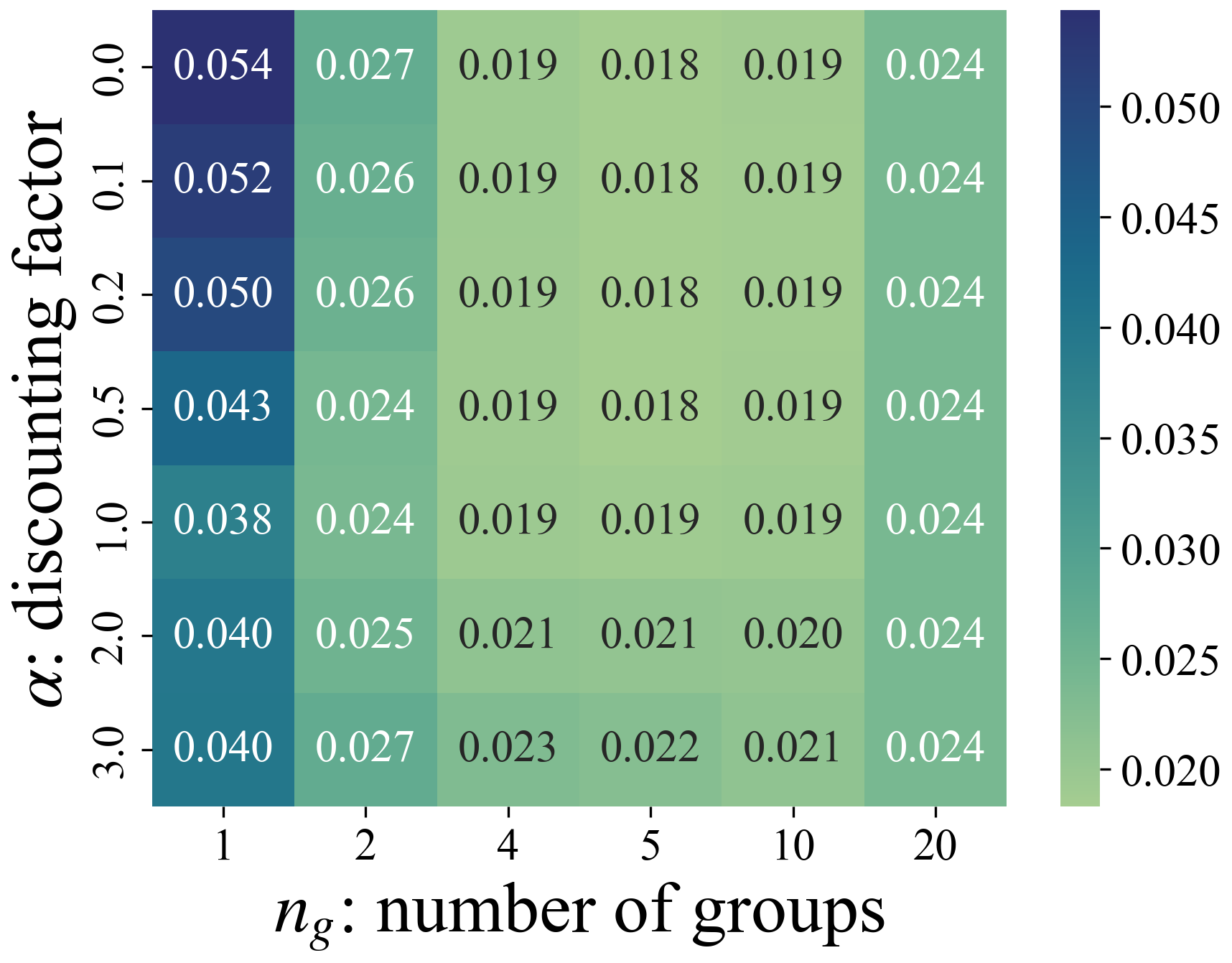}}
		\Description[Heat map showing the rank-discounted expected calibration error at top-N items.]{Heat map showing the rank-discounted expected calibration error at top-N items with color scale from 0.02 to 0.05 and the discounting factor from 3.0 to 0.0 on the Y axis against the number of groups from 1 to 20 on the X axis.}
		\caption{The sensitivities to the discounting factor $\alpha$ and the number of groups $n_g$. Calibration models are isotonic regression, Gamma calibration, isotonic regression, for MF, LightGCN, BPR, respectively.}
		\label{fig:sens_hp_additional}
	\end{center}
\end{figure}

In this section, we show several experimental results that additionally support the verification of proposed method.
We show the dependence on the number of recommendations for additional recommender models in Fig. \ref{fig:dep_nrec_additional}.
Fig. \ref{fig:sens_hp_additional} shows the sensitivity to hyperparameters for additional recommender models.

As a reference, we show the ranking-metrics and RMSE for each recommender model, that are independent from calibration,  in Table \ref{tab:eval_accuracy}.

\begin{table}[htbp]
	\caption{Evaluation of recommendation models}
	\label{tab:eval_accuracy}
	\begin{tabular}{lccc}
		\toprule
		& RMSE  & AUC & NDCG@20  \\
		\midrule
		itemKNN  &  1.017 $\pm$ 0.002 &  - & - \\
		MF  &  0.871 $\pm$ 0.001 &  - & - \\
		NCF   &  - &  0.787 $\pm$ 0.003 & 0.445 $\pm$ 0.007  \\
		LightGCN   &  - &  0.786 $\pm$ 0.002 & 0.417 $\pm$ 0.005  \\
		BPR  &  - &  0.801 $\pm$ 0.002 & 0.471 $\pm$ 0.004  \\
		\bottomrule
	\end{tabular}
\end{table}

\end{document}